\documentclass[
  aps,
  prb,
  reprint,
  amsmath,
  amssymb,
  superscriptaddress,
  floatfix,
  longbibliography,
]{revtex4-2}

\pdfoutput=1

\usepackage{graphicx}
\usepackage{physics}
\usepackage{amsmath}
\usepackage{amssymb}
\usepackage{amsbsy}
\usepackage{bm}
\usepackage{bbm}
\usepackage{float}
\usepackage{dsfont}
\usepackage{soul}
\usepackage{xcolor}
\usepackage{url}
\usepackage{hyperref}
\usepackage{booktabs}
\usepackage{tabularx}
\usepackage{threeparttable}
\usepackage{tikz}
\usepackage[normalem]{ulem} 

\hypersetup{
  colorlinks=true,
  citecolor=magenta,
  linkcolor=magenta,
  urlcolor=magenta
}

\DeclareUnicodeCharacter{2212}{-}

\begin{document}
\preprint{APS}
\title{Anisotropic g-tensor and strong spin-orbit interaction of holes in  hexagonal germanium nanowires}
 \author{Anirban Das}
 \affiliation{Department of Theoretical Physics, Institute of Physics, Budapest University of Technology and Economics, M\H{u}egyetem rkp.~3., H-1111 Budapest, Hungary}
 \author{Baksa Kolok}
\affiliation{HUN-REN-BME-BCE Quantum Technology Research Group, Budapest University of Technology and Economics, M\H{u}egyetem rkp.~3., H-1111 Budapest, Hungary}
 \affiliation{Department of Theoretical Physics, Institute of Physics, Budapest University of Technology and Economics, M\H{u}egyetem rkp.~3., H-1111 Budapest, Hungary}
\author{Dániel Varjas}
 \affiliation{Department of Theoretical Physics, Institute of Physics, Budapest University of Technology and Economics, M\H{u}egyetem rkp.~3., H-1111 Budapest, Hungary}
\affiliation{IFW Dresden and Würzburg-Dresden Cluster of Excellence ct.qmat, Helmholtzstrasse 20, 01069 Dresden, Germany}
   \author{András Pályi}
   \affiliation{HUN-REN-BME-BCE Quantum Technology Research Group, Budapest University of Technology and Economics, M\H{u}egyetem rkp.~3., H-1111 Budapest, Hungary}
 \affiliation{Department of Theoretical Physics, Institute of Physics, Budapest University of Technology and Economics, M\H{u}egyetem rkp.~3., H-1111 Budapest, Hungary}

\begin{abstract}
Hexagonal germanium combines direct-gap optical activity with the low-nuclear-spin environment of a group-IV material, making it a candidate platform for spin qubits and spin-photon interfaces. Using a ten-band $\mathbf{k}\cdot\mathbf{p}$ model, we study the response of confined hole states in hexagonal germanium nanowires to magnetic and electric fields. We show that, by symmetry, the top valence band of the bulk crystal has a vanishing $g$ factor in the plane perpendicular to the $c$ axis, in analogy with the in-plane $g$ factor of heavy holes in two-dimensional hole gases. In nanowires, confinement induces only small transverse $g$-tensor components, so the $g$ tensor remains strongly anisotropic. A transverse electric field strongly reduces the nanowire band gap, providing a gate-tunable optical signature and limiting the usable field range. The same field induces a Rashba spin-orbit interaction that is markedly stronger for fields perpendicular to the $c$ axis and depends nonlinearly on the field strength. In narrow wires at strong fields, the spin-orbit length reaches a few tens of nanometers, comparable to hole devices in cubic silicon. We identify the microscopic origin of the Rashba coupling as a second-order process mediated by a few intermediate subbands with distinct symmetry selection rules. Our results provide guidance for the design of electrically controlled hole-spin qubits in hexagonal germanium nanowires.

\end{abstract}

\maketitle


\section{Introduction}

The pursuit of scalable quantum computing has driven intensive exploration of semiconductor spin qubits, rooted in the foundational proposal by Loss and DiVincenzo to utilize electron spins in quantum dots as qubits \cite{loss1998quantum}. Early implementations in GaAs heterostructures demonstrated proof-of-concept spin control, but the strong hyperfine coupling to the host nuclear spin bath led to rapid inhomogeneous dephasing, with typical T$_2^\ast$ values of $\sim$ 10$~\mathrm{ns}$. This limitation motivated the subsequent development of group-IV platforms with reduced nuclear-spin noise \cite{merkulov2002electron,khaetskii2002electron,petta2005coherent, PhysRevB.74.195301,chekhovich2013nuclear,zaporski2023ideal}. 

Silicon offers intrinsically weak hyperfine coupling, which can be further suppressed by isotopic enrichment in $^{28}\mathrm{Si}$. This has enabled gate-defined quantum-dot electron spin qubits with inhomogeneous dephasing times reaching T$_2^\ast \sim$ 120$~\mu\mathrm{s}$, while echo and dynamical-decoupling protocols can extend coherence into the millisecond regime in optimized devices \cite{veldhorst2014addressable,kawakami2014electrical,yoneda2018quantum,burkard2023semiconductor}. These advances establish silicon as a leading platform for high-fidelity spin qubits. 

In parallel, germanium has emerged as a leading host for hole-spin qubits \cite{scappucci2021germanium}. The strong intrinsic spin--orbit interaction of holes enables fast, all-electrical spin manipulation without micromagnets \cite{watzinger2018germanium,fronthal2023hole}. In planar Ge/SiGe heterostructures, this has enabled fast single- and two-qubit gates and up to 18-qubit processors \cite{hendrickx2020fast,hendrickx2021four, dijkema2026simultaneous}, as well as hybrid devices with superconductors \cite{hendrickx2018gate, lakic2025quantum, valvo2026light}.

However, conventional cubic Si and Ge remain limited for optical interfacing because their indirect band gaps suppress efficient dipole-allowed band-edge transitions, hindering the direct integration of spin qubits with photonic quantum networks \cite{fadaly2020direct,deabreu2019characterization,kimble2008quantum,awschalom2018quantum}.

Recent breakthroughs in hexagonal Ge and SiGe alloys have begun to address this challenge. Unlike their cubic counterparts, hexagonal Ge and SiGe alloys can exhibit direct-bandgap optical emission with strong radiative recombination, enabling composition-tunable photoluminescence in the near- to mid-infrared range \cite{hauge2017single,cartoixa2017optical,fadaly2020direct,borlido2021pseudo,van2023polarized,van2024stimulated,peeters2024direct,broderick2026first}. This unique combination of direct-gap optical activity and a group-IV material base makes hexagonal SiGe a promising platform for integrated optoelectronic and potentially spin-photon applications \cite{berghuis2024low,van2024stimulated,peeters2024direct}. Furthermore, the hexagonal crystal symmetry modifies the electronic structure and valence-band mixing, which may enhance spin--orbit effects and enable all-electrical qubit control through electric-field-tunable $g$-factors and Rashba-type interactions \cite{winkler2003spin,crippa2018electrical,venitucci2018electrical,terrazos2021theory,liu2022emergent,pulcu2024multiband}.

Spin-orbit coupling and electric-field control of effective $g$-factors have been extensively investigated both theoretically and experimentally in conventional semiconductor nanowires and quantum wells, including III-V materials such as InAs, InSb, and GaAs, as well as cubic group-IV heterostructures based on Si and Ge \cite{nitta1997gate,nowack2007coherent,nadjperge2010spin,manchon2015rashba,kloeffel2018direct,kammhuber2017conductance,froning2021strong}. While these systems have served as important testbeds for spin-orbit physics, they do not simultaneously combine strong electric-field-tunable spin-orbit coupling, a group-IV material base, and efficient direct-gap optical activity, whereas III-V platforms are optically active but generally retain strong hyperfine coupling to host nuclear spins. This motivates the search for group-IV materials that can combine spin-qubit functionality with efficient optical interfaces.

Hexagonal Ge nanowires combine direct-gap optical activity with strong confinement-induced band mixing, offering a fundamentally different platform in which spin, orbital, and optical degrees of freedom can be coupled within a group-IV material system. Despite rapid progress in the growth and optoelectronic characterization of hexagonal Ge and SiGe nanostructures \cite{hauge2017single,cartoixa2017optical,fadaly2020direct,van2023polarized,peeters2024direct,broderick2026first}, their spin-qubit-relevant properties have only begun to be explored \cite{pulcu2024multiband,kolok2025acoustic}. In particular, the geometry- and electric-field-dependent effective $g$-tensor and Rashba spin-orbit interaction have not yet been quantitatively established for quantum-confined 2H-Ge nanowires. Beyond spin qubits, a quantitative understanding of these spin-dependent properties is also of crucial importance for engineering topological superconductivity and realizing Majorana bound states in semiconductor nanowire platforms \cite{lutchyn2010majorana,oreg2010helical, laubscher2024majorana, Maier2014majorana}. Here, we provide a theoretical analysis of these key spin-dependent properties, establishing a foundation for spin-qubit and spin-photon device design in this emerging material platform.

In this work, we develop a microscopic multiband theory of the magnetic- and electric-field response of confined hole states in 2H-Ge nanowires. Starting from a ten-band $\mathbf{k}\cdot\mathbf{p}$ Hamiltonian \cite{pulcu2024multiband}, we construct a real-space discretized nanowire model with open boundary conditions in the transverse plane and translational invariance along the nanowire axis. This framework allows us to analyze quantum-confinement-induced subband formation, compute the anisotropic effective $g$-tensor for longitudinal and transverse magnetic fields, and investigate the response of the top valence subband to transverse electric fields. We quantify the resulting Rashba spin-orbit interaction both from the electric-field-induced band splitting and through a Schrieffer-Wolff perturbative treatment, which further allows us to identify the parameter regimes where the perturbative description remains valid. 

Our main results are summarized as follows.
\begin{enumerate}
\item
\emph{Vanishing transverse $g$-factors.}
By symmetry arguments, we show that the top valence band of bulk hexagonal Ge has a single nonzero $g$-tensor component, along the $c$ axis: $|g_{zz}| \approx 18.08$, whereas the transverse $g$-factors vanish, i.e., $g_{xx} = g_{yy}  = 0$. This is analogous to the suppressed in-plane $g$-factor of heavy holes in two-dimensional hole gases in cubic materials \cite{winkler2003spin}. 
This analogy was missed in \cite{pulcu2024multiband} (as well as in our previous work \cite{kolok2025acoustic}), where the bulk transverse $ g$-factor was erroneously identified with the vacuum value of 2.

\item 
\emph{Highly anisotropic nanowire $g$-tensor.}
For hexagonal Ge nanowires, we find that the transverse confinement induces small transverse components $|g_{xx}|, |g_{yy}| \leq 1$ that tend to zero with increasing wire cross section, while $|g_{zz}|$ approaches its bulk value.
This implies that $|g_{xx}|, |g_{yy}| \ll |g_{zz}|$ holds in the nanowire's topmost valence subbands.

\item 
\emph{Electric-field-induced reduction of the fundamental gap.}
We identify a strong Stark shift of the topmost valence subband and the lowermost conduction subband of hexagonal Ge nanowires, caused by a transverse electric field. This Stark shift leads to a strong suppression of the fundamental gap for electric-field strengths of $\approx 10 \mathrm{V} / \mu\mathrm{m}$.
This effect provides a gate-tunable optical redshift and an upper bound on the usable electric-field strengths.

\item 
\emph{Rashba strength.}
We compute the strength $\alpha_R$ of the Rashba spin-orbit term as a function of electric-field strength and direction. 
For weak fields, we show that the Rashba effect is stronger (weaker) if the electric field is perpendicular to (parallel to) the $c$ axis.
At strong fields, we find a nonlinear dependence of $\alpha_R$ on the electric field, and conclude that it is maximized around $\alpha_R \approx 16 \, \mathrm{meV} \, \mathrm{nm}$ by using a narrow nanowire, a strong electric field of a few $\mathrm{V}/\mu\mathrm{m}$, and an electric field almost perpendicular to the $c$ axis. 
We also establish the electric-field range and momentum range where the often-used linear-in-$k_x$ Rashba model is valid. 

\item
\emph{Spin-orbit length.}
For strong electric fields, the spin-orbit length can be as low as $\approx 60 \, \mathrm{nm}$, comparable to spin-orbit length values reported for p-type planar MOS, FDSOI nanowire, and FinFET devices of cubic silicon, and slightly longer than cubic Ge/Si core/shell and Ge hut wire devices. 

\item
\emph{Microscopic origin of the Rashba effect.}
We establish the weak-field Rashba effect of the topmost valence subband as a second-order perturbative effect that is mediated by a few other subbands as intermediate virtual states, each of them coupled to the topmost valence subband by an electric-field matrix element and a $k_x$-linear matrix element.

\end{enumerate}

Our results reveal a pronounced intrinsic anisotropy in both the $g$-tensor and Rashba response, and provide quantitative guidance for electrically controlled hole-spin qubits in 2H-Ge nanostructures.

The paper is organized as follows. Section~II introduces the 2H-Ge nanowire model, including the ten-band $\mathbf{k}\cdot\mathbf{p}$ Hamiltonian, its real-space discretization, and the resulting confined valence subbands. Section~III investigates the magnetic-field response of the confined hole states and determines the effective $g$-tensor components for magnetic fields applied along the longitudinal and transverse directions. Section~IV examines the response to transverse electric fields. After introducing the electric-field Hamiltonian, we describe the quadratic Stark effect, derive a finite-field effective Hamiltonian for the top valence subband, and use it to analyze the Rashba splitting, its nonlinear field dependence, its angular anisotropy, and its dependence on the nanowire cross-sectional size. Section~V investigates the microscopic origin of the Rashba interaction by deriving the symmetry selection rules and identifying the dominant intermediate subbands. Section~VI compares the predicted Rashba coupling in 2H-Ge nanowires with that of other semiconductor nanowire platforms, and Section~VII summarizes our conclusions.

\section{Modeling of $\mathrm{\textbf{2H-Ge}}$ nanowires}

\begin{figure}[t]
    \centering
    \includegraphics[width=\columnwidth]{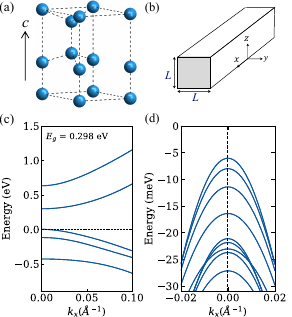}
    \caption{
\textbf{Crystal structure and electronic structure of hexagonal Ge nanowires.}
(a) Crystal structure of lonsdaleite (2H) Ge created in VESTA \cite{momma2011vesta}.
(b) Schematic of the nanowire geometry with square cross section of side length $L$.
(c) Bulk electronic band structure of 2H-Ge near $k_x=0$, showing a direct band gap of $E_g=0.298$ eV.
(d) Calculated valence subbands of a 2H-Ge nanowire oriented along the $x$ direction for $L=30$ nm.
}
    \label{unitcell}
\end{figure}

Motivated by experimentally realized hexagonal Si$_{1-x}$Ge$_x$ and Ge nanowires~\cite{lamon2025dimension,lamon2026hexagonal,li2022hexagonal}, we consider a 2H-Ge nanowire with a square cross section and its longitudinal axis oriented perpendicular to the crystal $c$ axis. We choose the nanowire axis as the $x$ direction, corresponding to the $[10\bar{1}0]$ crystallographic direction, and align the $z$ axis with the crystal $c$ axis. The carriers are therefore confined in the transverse $y$--$z$ plane, as illustrated schematically in Fig.~\ref{unitcell}(b). The square cross section has side length $L$; we consider the range $20 \, \mathrm{nm} \leq L \leq 60 \, \mathrm{nm}$ in this work. Unless stated otherwise, we use $L=30~\mathrm{nm}$ as a representative value.

To describe its low-energy electronic structure near the Brillouin-zone center, we employ the ten-band $\mathbf{k}\cdot\mathbf{p}$ Hamiltonian developed for 2H-Ge \cite{pulcu2024multiband,Chunag96}. The full matrix representation of the Hamiltonian is given in Eq.~\eqref{eq:hamiltonian}.
Note that $H_{\mathbf{k}\cdot\mathbf{p}}$ in Eq.~\eqref{eq:hamiltonian} differs from the explicit matrix given in Ref.~\cite{pulcu2024multiband}. There, the linear ($P_1$, $P_2$) and second-order ($A_6$) couplings of the valence-band states are written in inconsistent phase conventions. As a result, the Hamiltonian is invariant only under the $C_{6h}$ subgroup of $D_{6h}$: the vertical mirror planes and the in-plane $C_2$ axes are broken. We instead use a form in which all terms follow from a single basis and which respects the full $D_{6h}$ symmetry and time reversal (see Appendix~\ref{app:hamiltonian} for detailed description).

This model reproduces the density-functional-theory band structure near the $\Gamma$ point and yields a direct band gap of $0.298~\mathrm{eV}$ \cite{Claudia2019accurate}. The corresponding bulk band structure is shown in Fig.~\ref{unitcell}(c). In the absence of external magnetic and electric fields, all bands remain spin degenerate because of the combined presence of time-reversal and inversion symmetry.

To construct the confined nanowire model, we regularize the continuum Hamiltonian on a real-space lattice. The linear and quadratic momentum operators are replaced by their lattice-regularized forms,
\begin{align}
    k_i &\rightarrow \frac{1}{\Delta}\sin(k_i\Delta),\\
    k_i^2 &\rightarrow
    \frac{2}{\Delta^2}\left[1-\cos(k_i\Delta)\right],
\end{align}
where $i\in\{x,y,z\}$ and $\Delta$ is the discretization length. Translational invariance is retained along the nanowire axis, such that $k_x$ remains a good quantum number, whereas the transverse $y$ and $z$ directions are represented on a real-space grid with open boundary conditions.

Equivalently, the same lattice Hamiltonian can be obtained by replacing $k_i\rightarrow-i\partial_i$, for $i \in \{y,z\}$, in the continuum Hamiltonian and discretizing the resulting differential operators on a grid of spacing $\Delta$ using lowest-order symmetric finite differences. Keeping the momentum variable $k_x$ along the translationally invariant $x$ direction then produces a tight-binding Hamiltonian in the mixed $(k_x,y,z)$ representation.

The resulting nanowire Hamiltonian can be written as
\begin{align}
H_0(k_x)
=
\sum_{i,j=1}^{N}
\sum_{a,b=1}^{10}
t^{ij,k_x}_{ab}\,
c^\dagger_{i a k_x}
c_{j b k_x},
\label{eq:H0}
\end{align}
where $i$ and $j$ label lattice sites in the transverse $y$-$z$ cross section and $N$ denotes the total number of transverse sites. The indices $a$ and $b$ label the ten internal orbital and spin degrees of freedom. The operators $c^\dagger_{i a k_x}$ and $c_{i a k_x}$ create and annihilate, respectively, a particle in internal state $a$ at transverse site $i$ with wave vector $k_x$. The matrix elements $t^{ij,k_x}_{ab}$ contain both the on-site terms and the hopping amplitudes between transverse sites $i$ and $j$ at fixed $k_x$. Diagonalizing Eq.~\eqref{eq:H0} yields the confined nanowire subband dispersion.

Throughout this work, we distinguish bulk bands from confined nanowire subbands using the following notation. Bulk conduction bands are denoted by $\mathrm{CB}_1,\mathrm{CB}_2$, where $\mathrm{CB}_1$ is the lowest bulk conduction band. Bulk valence bands are denoted by $\mathrm{VB}_1,\mathrm{VB}_2,\mathrm{VB}_3$, where $\mathrm{VB}_1$ is the highest bulk valence band at $\Gamma$, and the index increases toward lower-energy valence bands. Confined nanowire valence subbands are denoted by
$\mathrm{vb}_0,\mathrm{vb}_1,\mathrm{vb}_2,\ldots$, where $\mathrm{vb}_0$ is the highest confined valence subband. Thus, a
statement such as ''$\mathrm{vb}_0$ has predominantly $\mathrm{VB}_1$ character'' means that the highest confined nanowire
valence subband is dominated by the top bulk-valence-band component.

As a representative example, Fig.~\ref{unitcell}(d) shows the confined valence-subband dispersion near $k_x=0$ for a square nanowire with $L=30~\mathrm{nm}$.

\section{Magnetic-field Response}
\label{sec:magnetic_response}

To investigate the response of 2H-Ge nanowires to an external magnetic field, we analyze the Zeeman splitting of the subbands derived from the top valence band. The magnetic response provides direct access to the effective $g$ tensor, which is a central experimentally accessible quantity in magneto-transport and spectroscopic measurements and is of direct relevance to spin-qubit implementations.

The magnetic field $\mathbf{B}$ enters the nanowire Hamiltonian through two distinct contributions. The orbital coupling is incorporated via the Peierls substitution \cite{peierls1933,hofstadter1976}
\begin{equation}
\mathbf{k} \rightarrow \mathbf{k} - \frac{e}{\hbar}\mathbf{A}(\mathbf{r}),
\label{eq:peierls}
\end{equation}
where $\mathbf{A}(\mathbf{r})$ is the magnetic vector potential. Here, $e>0$ denotes the elementary charge, and the signs are chosen according to the electron-band Hamiltonian convention used throughout this work. This substitution modifies the hopping amplitudes in the discretized Hamiltonian, giving rise to an \textit{orbital Zeeman effect} whose form depends explicitly on the magnetic-field orientation and the nanowire geometry. The explicit expressions for the modified hopping terms are derived in Appendix~\ref{app:peierls}.

\begin{figure}[t]
 \centering
 \includegraphics[width=\linewidth]{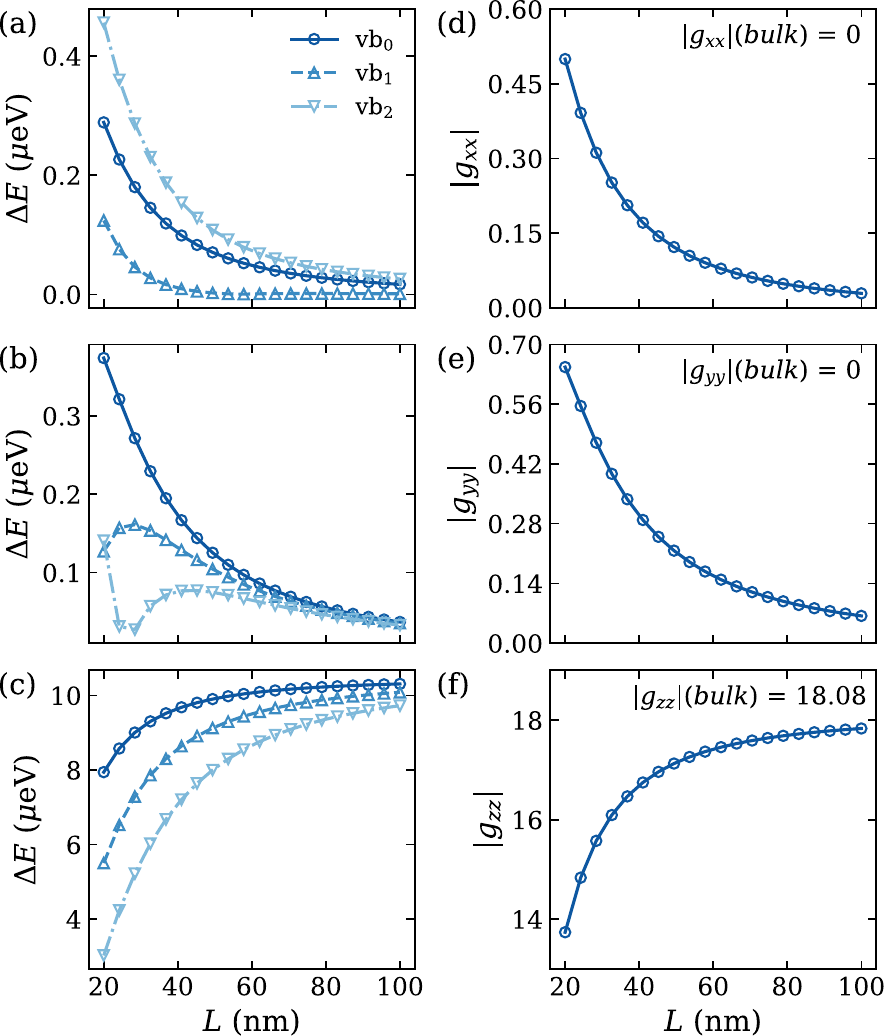}
\caption{\textbf{Zeeman splitting and effective $g$-factors.}
(a)-(c) Zeeman splitting energy $\Delta E$ of the top three valence subbands ($\mathrm{vb}_0$, $\mathrm{vb}_1$, and $\mathrm{vb}_2$) as a function of the nanowire cross-sectional side length $L$, for a magnetic field applied along the $x$, $y$, and $z$ directions, respectively. The magnetic field strength is $B=0.01$ T. (d)-(f) Corresponding absolute effective $g$-factors of the top valence subband.}
\label{zeeman}
\end{figure}

In addition, the general Hamiltonian describing the \textit{spin Zeeman effect} is given by
\begin{equation}
H_{\mathrm{spin}} = g_0 \mu_B \, \boldsymbol{S} \cdot \mathbf{B},
\label{eq:spin_zeeman}
\end{equation}
where $g_0 \approx 2$ denotes the free-electron $g$-factor, $\mu_B$ is the Bohr magneton, and $\boldsymbol{S}$ represents the spin vector in $\hbar$ units.
In the basis of the 10-band Hamiltonian in Eq.~\eqref{eq:hamiltonian}, the term $\boldsymbol{S} \cdot \mathbf{B}$ can be written explicitly as
\begin{align}
    \boldsymbol{S} \cdot \mathbf{B} &= \frac{1}{2}[B_x (\sigma_x \otimes \mathbb{M}_5) + B_y (\sigma_y \otimes \mathbb{M}_5)\nonumber
    \\ &\quad + B_z(\sigma_z\otimes\mathbb{I}_5)],
\end{align}
where $\mathbb{I}_5$ is the $5\times 5$ identity matrix and $\mathbb{M}_5$ has the form,
\begin{align}
        \mathbb{M}_5 = \mqty(1 & 0 & 0& 0& 0\\ 
                                0 & 1& 0& 0& 0\\
                                0& 0& 0& 1& 0\\
                                0& 0& 1& 0& 0\\
                                0& 0& 0& 0& 1).
    \end{align}

The particular form of $\mathbb{M}_5$ originates from the ordering of the symmetry-adapted basis used for the 10-band $\mathbf{k}\cdot\mathbf{p}$ Hamiltonian, given in Table~\ref{tab:basis}. 
The reason for this is that the order of time-reversed partners is maintained, but the orbital parts of the third and fourth basis states are also exchanged by time reversal. Hence a spin Zeeman term acting only on the spin obtains this form. A result of this is the vanishing perpendicular $g$-factors $|g_{xx}|=|g_{yy}|=0$ in the top bulk valence band, similar to heavy hole -- light hole systems; for a proof, see Appendix~\ref{app:bulk_xy_gfactor}.
We note that Table~VI of Ref.~\cite{pulcu2024multiband}, and Table~I. of Ref.~\cite {kolok2025acoustic} incorrectly report $|g_{xx}|=|g_{yy}|=2.0$.
We found that the reported value of $g_{zz}$ was also inaccurate; a computation of the bulk $g_{zz}$ is presented in Appendix~\ref{app:bulkgfactor}.

For hole states in low-symmetry and strongly confined systems, spin and orbital degrees of freedom are intrinsically coupled, such that the experimentally observed Zeeman splitting reflects the combined effect of both contributions \cite{CsontosPRB09, froning2021strong, zhang2021anisotropic}.

Throughout this section, we consider magnetic fields applied along the nanowire axis ($x$ direction) and along the two transverse directions ($y$ and $z$). In the presence of a magnetic field, the Kramers degeneracy of the valence subbands is lifted. For a given field orientation $i \in \{x,y,z\}$, we define the Zeeman splitting of the Kramers doublet as
\begin{equation}
\Delta E(B_i) = |E_1(B_i)-E_2(B_i)|,
\label{eq:zeeman_splitting}
\end{equation}
where $E_1$ and $E_2$ are the two field-split partners of the Kramers doublet at $k_x=0$ for a field along direction $i$.
The corresponding effective $g$ factor is extracted from the linear response in the weak-field limit,
\begin{equation}
g_{ii} = \frac{\Delta E}{\mu_B |B_i|}
\label{eq:gi_definition}
\end{equation}
This definition directly corresponds to the effective $g$ factors measured experimentally and naturally incorporates both orbital and spin contributions.

Figures \ref{zeeman}(a)–\ref{zeeman}(c) show the Zeeman splitting of the top three valence subbands as a function of the nanowire cross-sectional size for a fixed magnetic-field magnitude. In the strong-confinement regime, the splittings differ substantially between subbands, reflecting confinement-induced band mixing. As the cross-sectional size increases, the splittings gradually approach their bulk-like behavior.

The effective $g$ factors extracted for the topmost valence subband are shown in Figs.~\ref{zeeman}(d)--\ref{zeeman}(f). We find a pronounced anisotropy of the $g$ tensor. For magnetic fields applied along the nanowire axis ($x$-direction) and the transverse $y$ direction, the magnitudes of the corresponding components $|g_{xx}|$ and $|g_{yy}|$ decrease with increasing cross-sectional size and approach values close to zero in the bulk limit. In contrast, for magnetic fields applied along the crystal $c$ axis ($z$ direction), $|g_{zz}|$ remains large and saturates near the bulk value of approximately $18.08$. This behavior reflects the strong anisotropy of the valence band in hexagonal Ge.

The anisotropic $g$ tensor reported here has direct implications for spin-qubit implementations in hexagonal Ge nanowires. The large $g_{zz}$ component, corresponding to magnetic fields applied along the crystal $c$ axis, enables sizable Zeeman splittings at moderate magnetic fields. In contrast, fields applied along the nanowire axis or along the orthogonal transverse direction produce significantly smaller splittings. This strong anisotropy provides practical guidance for choosing the magnetic-field orientation in spin-qubit experiments, particularly when a sizable and controllable Zeeman splitting is required. These features are especially relevant for experiments that employ tilted magnetic fields to extract the full $g$ tensor and optimize qubit operation and coherence.

\section{Electric-field Response}
\label{sec:finite_field}

In this section, we analyze the effect of a homogeneous electric field $\boldsymbol{\mathcal{E}}$ transverse to the nanowire axis.
Such electric fields are often present in gated devices: for a nanowire lying on a horizontal substrate, bottom and top gates (side gates) provide an approximately vertical (horizontal) electric field.
To describe the effects of such electric fields, we first generalize our discretized model Hamiltonian for the hexagonal germanium nanowire, Eq.~\eqref{eq:H0}, to include the electric field in Sec.~\ref{sec:electrichamiltonian}.
In Sec.~\ref{subsec:stark_shift_sub}, we identify and quantitatively describe the quadratic Stark shift of the electronic energies, which reduces the fundamental gap of the nanowire and sets the electric-field range considered in the rest of this section.
The remaining subsections address the Rashba-type spin splitting induced by the electric field.
In Sec.~\ref{subsec:finite_field_rashba}, we derive a finite-field effective Hamiltonian for the top valence subband, define the Rashba coefficient, and show that the splitting is strongly anisotropic.
In Sec.~\ref{subsec:nonlinear_rashba_modulation}, we examine the nonlinear dependence of the Rashba coefficient on the strength and orientation of the field.
In Sec.~\ref{subsec:linear-in-kx_validity}, we determine the range of validity of the linear-in-$k_x$ Rashba model for gate-defined quantum dots.
Finally, in Sec.~\ref{subsec:rashba_size_dependence}, we map the joint dependence of the Rashba coefficient on field strength and cross-section size.

\subsection{Electric-field Hamiltonian}
\label{sec:electrichamiltonian}

 In the discretized real-space representation, the position operator is diagonal in both the transverse-site index and the internal band index. Using the same index structure as the field-free nanowire Hamiltonian, the electric-field Hamiltonian at fixed $k_x$ can be written as
\begin{align}
H_{\mathcal{E}}
=
\sum_{i,j=1}^{N}
\sum_{a,b=1}^{10}
\left[
e\left(
\boldsymbol{\mathcal{E}}\cdot\mathbf{r}_i
\right)
\delta_{ij}\delta_{ab}
\right]
c^\dagger_{ia k_x}c_{jb k_x}.
\label{eq:electric_field_matrix_form}
\end{align}
Here, $\mathbf{r}_i=(y_i,z_i)$ denotes the position of the $i$'th transverse grid site. Thus, the electric field introduces a position-dependent on-site potential that is diagonal in the internal basis and does not couple different transverse sites.

The full Hamiltonian then reads as
\begin{align}\label{eq:EHamiltonian}
    H(\boldsymbol{\mathcal{E}}, k_x) = H_0(k_x) + H_{\mathcal{E}}.
\end{align}
Figs.~\ref{fig:stark_effect}(a) and \ref{fig:stark_effect}(b) show the $k_x$-dependent energy levels of the topmost valence subband, derived from the Hamiltonian in Eq.~\eqref{eq:EHamiltonian}.
Different colors represent dispersion relations for different electric-field strengths. The electric field is applied along the $y$ axis for panel (a) and along the $z$ axis for panel (b). With increasing field strength, an effect that is clearly observed in panels (a) and (b) is the Stark shift: the subbands shift upwards in energy. Another important observation is the Rashba-type spin splitting of the subbands, which arises at finite electric field and finite momentum $k_x > 0$; this is well observable in panel (a), but invisibly small in panel (b). In what follows, we analyze these effects in detail.

\subsection{Quadratic Stark effect}
\label{subsec:stark_shift_sub}

\begin{figure}[t]
    \centering
    \includegraphics[width=\columnwidth]{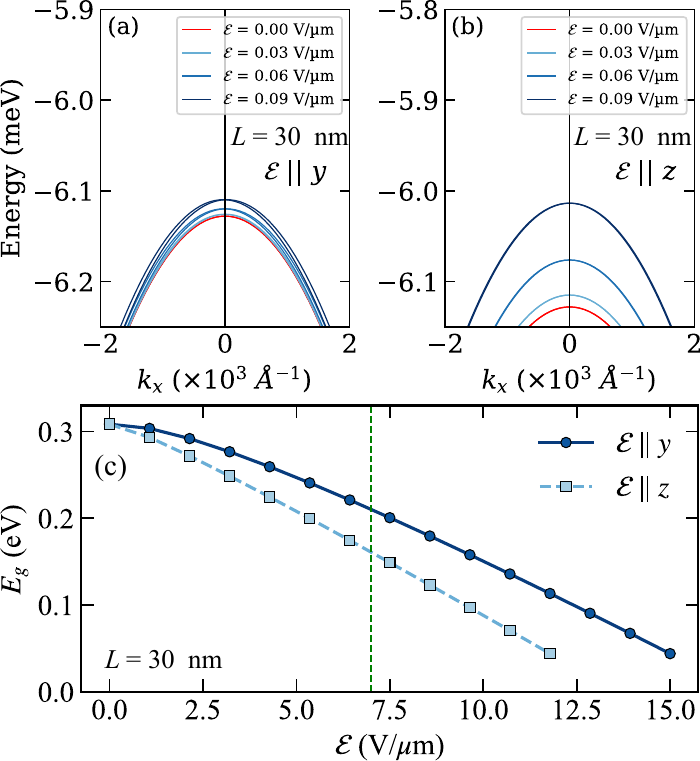}
    \caption{
    (a),(b) Dispersion of the top valence subband under transverse electric fields applied along the $y$ and $z$ directions,
    respectively. The different curves correspond to electric-field strengths $\mathcal{E}=0.00$, $0.03$, $0.06$, and
    $0.09~\mathrm{V/\mu m}$. The black vertical lines mark $k_x=0$. (c) Fundamental band gap $E_g$ as a function of electric-field strength, illustrating the quadratic Stark effect. The blue circles connected by the dark blue solid line and the light-blue squares connected by the dashed line correspond to electric fields applied along the $y$ and $z$ directions, respectively. The green dashed line represents the maximum electric field limit chosen for the rest of the calculations in Secs.~\ref{subsec:finite_field_rashba} and \ref{subsec:nonlinear_rashba_modulation}. For all panels, the side length of the square cross section of the nanowire is $L=30~\mathrm{nm}$.
    }
    \label{fig:stark_effect}
\end{figure}

The upward shift is distinct from the Rashba splitting and originates from the quadratic Stark effect~\cite{miller84,kuo2005strong}. It is strongly anisotropic: the shift produced by $\mathcal{E}_z$ (Fig.~\ref{fig:stark_effect}(b)) is substantially larger than that produced by $\mathcal{E}_y$
(Fig.~\ref{fig:stark_effect}(a)). 

This behavior can be understood from the effective-mass anisotropy, with the bulk effective masses $m_z^*=-0.524\,m_\mathrm{e}$ and $m_y^*=-0.0672\,m_\mathrm{e}$, where $m_\mathrm{e}$ is the free-electron mass.
(We express these effective masses from the bulk $\boldsymbol{\mathrm{k}}\cdot\boldsymbol{\mathrm{p}}$ Hamiltonian parameters in Appendix~\ref{app:effmass}.)
In the single-band hard-wall approximation at fixed cross-sectional size, the quadratic Stark shift is proportional to the effective mass. (This is revealed by an analytical perturbative evaluation of the Stark shift, see Appendix \ref{app:quadratic_stark_effect}.)
Because $|m_z^*|$ is much larger than $|m_y^*|$, the confinement-induced level spacing is smaller for modes associated with motion along $z$. 
A $z$-directed field therefore more efficiently mixes the top valence subband with excited transverse modes, resulting in a larger upward Stark shift.

The electric field induces a Stark shift also for the lowermost conduction band, and these together lead to a change of the nanowire band gap. Therefore, the electric field induces an experimentally relevant optical signature. 
Our numerical results show that the valence-band edge shifts upward and the conduction-band edge shifts downward, thereby reducing the fundamental gap of the nanowire. Fig.~\ref{fig:stark_effect}(c) shows this gap reduction for both field orientations. The corresponding transition energy is therefore expected to exhibit a redshift with increasing field strength. At sufficiently large fields, the electrostatic potential-drop energy across the nanowire, $e\mathcal E_iL$, becomes comparable to the band-gap energy, and the calculated gap approaches closure near $\mathcal E_y=15.5~\mathrm{V/\mu m}$ and $\mathcal E_z=12.5~\mathrm{V/\mu m}$. To remain well within the gapped regime for both field orientations, we therefore restrict the field magnitude to $\mathcal E\leq7~\mathrm{V/\mu m}$ in the calculations presented in Secs.~\ref{subsec:finite_field_rashba} and \ref{subsec:nonlinear_rashba_modulation}. This cutoff is indicated by the green dashed line in Fig.~\ref{fig:stark_effect}(c).

\subsection{Finite-field effective Hamiltonian and Rashba splitting}
\label{subsec:finite_field_rashba}

To investigate the Rashba splitting, we apply a Schrieffer--Wolff transformation to find the effective two-by-two, $ k_x$-dependent Hamiltonian for the topmost valence subband in the presence of the electric field.
The unperturbed Hamiltonian is defined as
\begin{equation} \label{eq:unperturbedHE}
    \widetilde{H_0}(\boldsymbol{\mathcal{E}}) = H(\boldsymbol{\mathcal{E}}, k_x = 0).
\end{equation} 
The remaining part of the Hamiltonian is taken into account up to second order in $k_x$; we will consider this as the perturbation:
\begin{equation}
 W = k_xV_1
    +k_x^2V_2.
\end{equation}
Here we introduced the notations
\begin{align}
    V_1
    &=
    \left.
    \frac{\partial H_0(k_x)}{\partial k_x}
    \right|_{k_x=0},\label{V1}\\
    V_2
    &=
    \left.
    \frac{1}{2}\frac{\partial^2H_0(k_x)}{\partial k_x^2}
    \right|_{k_x=0}.
    \label{V2}
\end{align}

The term linear in $k_x$ generates the leading Rashba splitting, while the quadratic term is required to reproduce the leading curvature of the subband dispersion.

Using the Schrieffer-Wolff transformation as implemented in the \textsc{Pymablock} Python package~\cite{10.21468/SciPostPhysCodeb.50-r2.1}, we obtain the $2\times2$ effective Hamiltonian on the subspace spanned by the Kramers pair of the topmost valence subband, computed from the unperturbed Hamiltonian in 
Eq.~\eqref{eq:unperturbedHE}. Taking $k_xV_1$ and $k_x^2V_2$ as the two perturbations, the effective Hamiltonian up to second order in $k_x$ is
\begin{align}
    H_{\mathrm{eff}}(\boldsymbol{\mathcal{E}},k_x)
    &=
    H_{00}(\boldsymbol{\mathcal{E}})
    +H_{10}(\boldsymbol{\mathcal{E}})k_x
    \nonumber\\
    &\quad+
    \left[
    H_{20}(\boldsymbol{\mathcal{E}})+H_{01}(\boldsymbol{\mathcal{E}})
    \right]k_x^2.
    \label{eq:effective_dispersion_kx}
\end{align}
Here, $H_{00}$ is the projection of the finite-field reference Hamiltonian $\widetilde H_0(\boldsymbol{\mathcal{E}})$ in Eq.~\eqref{eq:unperturbedHE} onto its Kramers doublet at $k_x=0$. 
The matrix $H_{10}$ is generated via first-order perturbation theory in $k_xV_1$, and determines the leading Rashba splitting.
As for the quadratic in $k_x$ terms, $H_{20}$ is obtained from second-order perturbation theory in $k_xV_1$, whereas $H_{01}$ is obtained from first-order perturbation theory in $k_x^2V_2$. 
Since time reversal acts in a spinful way on the Kramers pair, while leaving $k_x^2$ unchanged, the $H_{20}+H_{01}$ term cannot lift the Kramers degeneracy, hence is proportional to the identity.

The Rashba coefficient, which describes the energy splitting of the subband up to first order in $k_x$, is obtained from the term $H_{10}(\boldsymbol{\mathcal{E}}) k_x$.
The energy splitting due to this term is expressed from the two eigenvalues $\lambda_1(\boldsymbol{\mathcal{E}}) < \lambda_2(\boldsymbol{\mathcal{E}})$ of $H_{10}(\boldsymbol{\mathcal{E}})$ as 
\begin{equation}
    \label{eq:linearenergysplitting}
    \Delta E(\boldsymbol{\mathcal{E}}) =
    (\lambda_2(\boldsymbol{\mathcal{E}})-\lambda_1(\boldsymbol{\mathcal{E}})) |k_x| = 
    2 \alpha_R(\boldsymbol{\mathcal{E}}) |k_x|,
\end{equation}
where we define the Rashba coefficient as
\begin{equation}
\label{eq:rashbacoefficientdefinition}
    \alpha_R(\boldsymbol{\mathcal{E}}) = \frac{\lambda_2(\boldsymbol{\mathcal{E}})-\lambda_1(\boldsymbol{\mathcal{E}})}{2}.
\end{equation}
Note that the Rashba coefficient $\alpha_R$ has the dimension $\mathrm{energy} \times \mathrm{length}$.

\begin{figure}[t]
    \centering
    \includegraphics[width=\columnwidth]{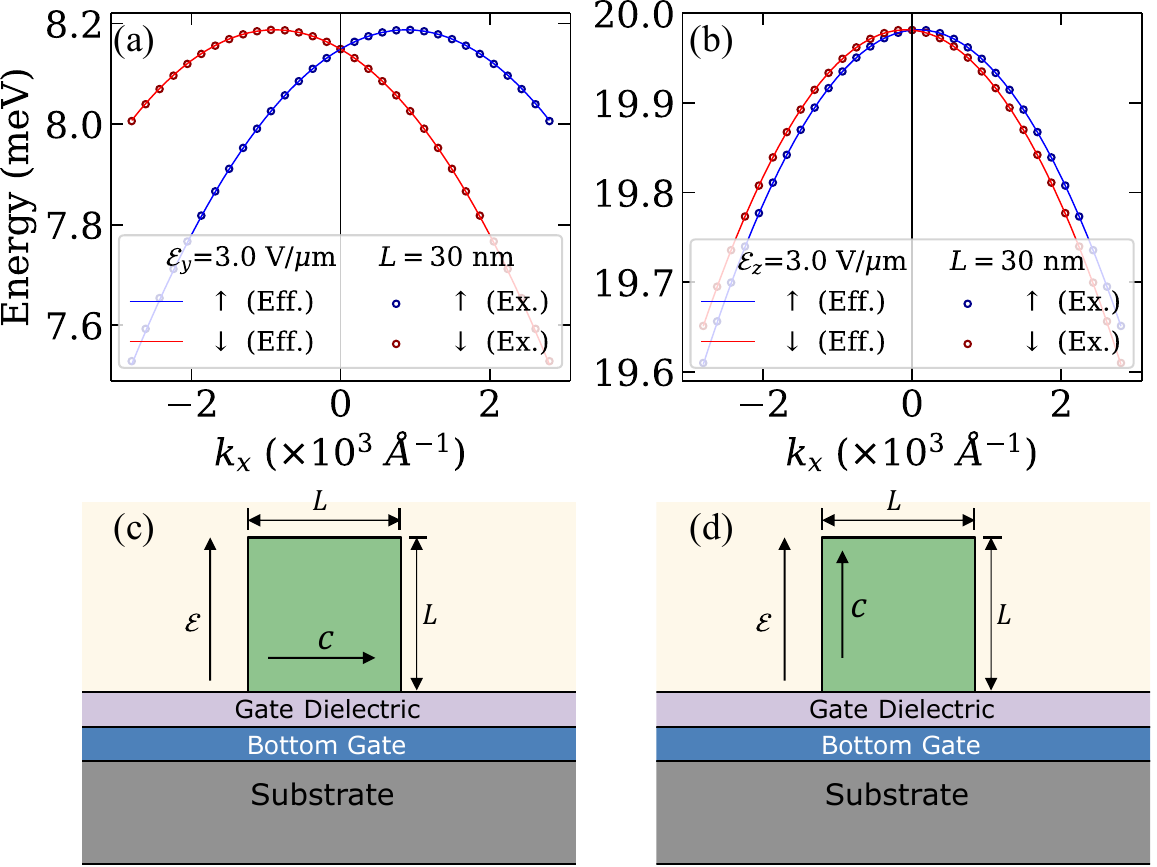}
    \caption{Highly anisotropic Rashba splitting in hexagonal germanium nanowires.
    (a,b) Momentum-dependent Rashba splitting of the topmost valence subband for
    (a) $\mathcal{E}_y=3.0~\mathrm{V/\mu m}$ and
    (b) $\mathcal{E}_z=3.0~\mathrm{V/\mu m}$ in a square nanowire with cross-sectional side length $L=30~\mathrm{nm}$. Dots show the dispersions obtained by exact diagonalization of the full Hamiltonian (Ex.), while the solid lines show the results obtained from the effective Hamiltonian through second order in $k_x$ (Eff.). Blue and red denote the two pseudospin branches that evolve from the Kramers doublet at $k_x=0$. 
    Arrows in the legend are used as labels only; they do not correspond to physical spin directions.
    Panels (c) and (d) illustrate bottom-gated device configurations in which the crystal $c$ axis is parallel and perpendicular to the gate plane, respectively. The vertical arrows indicate the approximate direction of the gate-induced electric field.
    }
    \label{fig:rashba_linear_kx}
\end{figure}

Fig.~\ref{fig:rashba_linear_kx}(a) and
\ref{fig:rashba_linear_kx}(b) show the momentum-dependent energy splitting for fields applied along $y$ and $z$, respectively, at the common field strength $3.0~\mathrm{V/\mu m}$. 
Both panels Fig.~\ref{fig:rashba_linear_kx}(a) and (b) compare the numerically exact dispersions (dots) with the perturbative dispersions (solid); the comparison shows negligible difference in both panels.
The two branches remain degenerate at $k_x=0$ because time-reversal symmetry is preserved. At finite momentum, the electric field breaks inversion symmetry and produces a Rashba-type splitting. 
The splitting generated by $\mathcal{E}_y$ (Fig.~\ref{fig:rashba_linear_kx}(a)) is substantially larger than that generated by $\mathcal{E}_z$ 
(Fig.~\ref{fig:rashba_linear_kx}(b))
at the same field strength, demonstrating the strong directional anisotropy of the Rashba interaction.

This anisotropy provides direct guidance to design the device geometry and nanowire orientation if the Rashba strength needs to be optimized.
Consider, for example, bottom-gated devices as shown in Fig.~\ref{fig:rashba_linear_kx}(c) and (d), where the gate-induced electric field is mostly vertical. 
To maximize the Rashba strength in such a device, the configuration in Fig.~\ref{fig:rashba_linear_kx}(c), where the crystal $c$ axis lies parallel to the gate, is preferred: in that case, the vertical electric field creates a much stronger Rashba splitting than in the configuration of Fig.~\ref{fig:rashba_linear_kx}(d).

\subsection{Nonlinear dependence of the Rashba coefficient on the electric-field strength}
\label{subsec:nonlinear_rashba_modulation}

For a weak electric field, the Rashba coefficient $\alpha_R$ defined in Eq.~\eqref{eq:rashbacoefficientdefinition} is a linear function of the electric-field strength. Here, we describe the situation when the electric field is strong enough that this linear relation breaks down.
We observe a nonlinear dependence of the Rashba coefficient on the field strength. Our results imply that for a strong electric field with a fixed strength, the Rashba coefficient is maximized if the field is slightly tilted with respect to the $y$ axis, and the optimal tilt angle depends on the field strength.

\begin{figure}[t]
    \centering
    \includegraphics[width=\columnwidth]{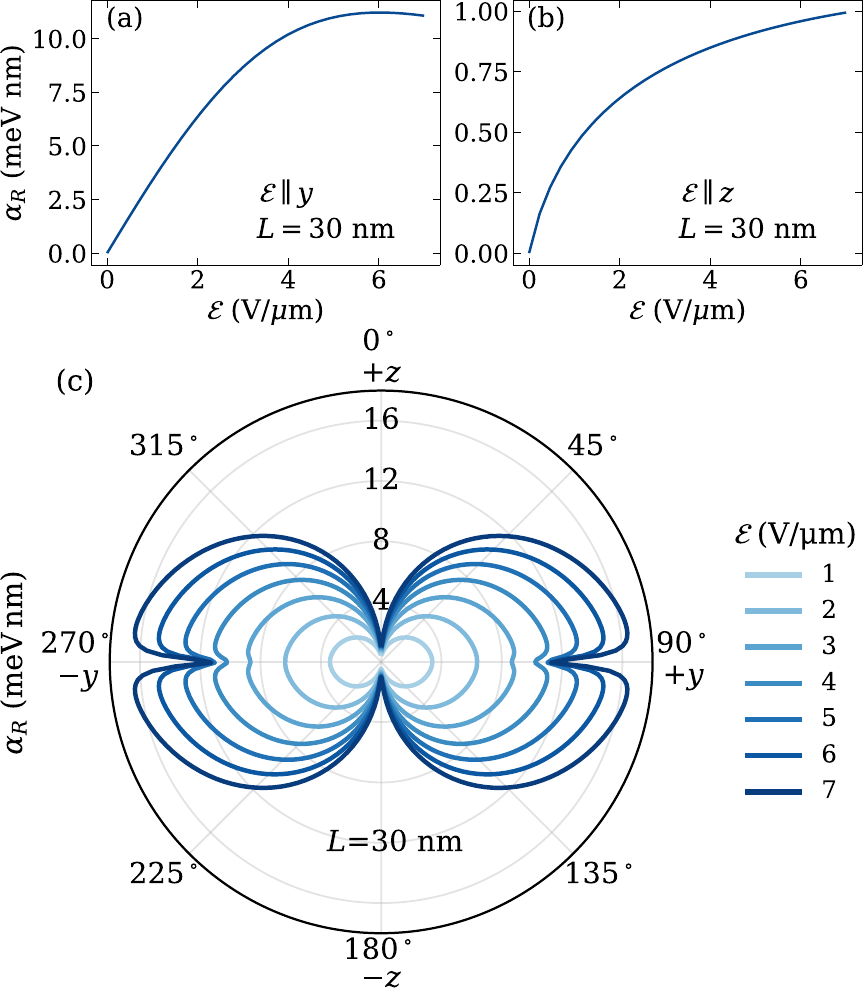}
    \caption{
    Electric-field dependence and anisotropy of the effective Rashba coefficient for a hexagonal germanium nanowire. 
    (a),(b) Variation of $\alpha_R$ as a function of electric fields applied along $y$ and $z$, respectively. (c) Polar representation of $\alpha_R$ for a field rotated in the $yz$ plane at fixed magnitudes in the range
    $1\, \mathrm{V}/\mu\mathrm{m} \leq
    \mathcal{E} \leq
    7\, \mathrm{V}/\mu\mathrm{m}$.
    The angles $\theta=0^\circ$ and $90^\circ$ correspond to  $\boldsymbol{\mathcal{E}}\parallel+z$ and $\boldsymbol{\mathcal{E}}\parallel+y$, respectively. Progressively darker blue curves denote increasing field strength.
    For all panels, the side length of the square cross section of the nanowire is $L=30~\mathrm{nm}$.
    }
    \label{fig:Fig-5}
\end{figure}

Our numerical results shown in Fig.~\ref{fig:Fig-5} indicate the nonlinear dependence of the Rashba coefficient on the electric-field strength, as well as an interesting type of anisotropy with respect to the electric-field orientation. Fig.~\ref{fig:Fig-5}(a) shows that for $\boldsymbol{\mathcal{E}}\parallel y$, the Rashba coefficient $\alpha_R$ increases linearly and rapidly with increasing $\boldsymbol{\mathcal{E}}$, reaches a maximum near $\mathcal{E}=6~\mathrm{V/\mu m}$, and then decreases weakly; i.e., the dependence is nonlinear and non-monotonic. Fig.~\ref{fig:Fig-5}(b) shows that for $\boldsymbol{\mathcal{E}}\parallel z$, the Rashba coefficient $\alpha_R$ is a nonlinear function of the electric field, although it is monotonic in the considered field range. The maximum value of the Rashba strength for $\boldsymbol{\mathcal{E}} \parallel y$ is approximately ten times greater than for $\boldsymbol{\mathcal{E}} \parallel z$. 

Fig.~\ref{fig:Fig-5}(c) shows the angular dependence of the Rashba strength. 
Each colored line corresponds to a fixed electric-field strength whose orientation is varied in the $yz$ plane perpendicular to the nanowire axis.
For weak fields, the Rashba coefficient is maximal along $\pm y$, and minimal along $\pm z$.  For stronger fields, however, the $\pm y$ directions become local minima, and the Rashba coefficient is maximized at a small tilt angle of the electric field with respect to the $y$ direction. These results also imply that the Rashba coefficient might change significantly due to a small change of direction of the electric field.

\subsection{Validity of the linear-in-$k_x$ Rashba model}
\label{subsec:linear-in-kx_validity}

The Rashba effect is an important ingredient in the functionality of spin qubits formed in electrically defined quantum dots.
On the positive side, it serves as a mechanism for qubit control via electrically driven spin resonance. 
On the negative side, it leads to information loss for such quantum-dot-based spin qubits in the presence of electric noise. 
In any case, the linear Rashba model we describe above, where the term $H_{10}(\boldsymbol{\mathcal{E}})$ of the effective Hamiltonian $H_\mathrm{eff}$ accounts for spin splitting (see Eq.~\eqref{eq:effective_dispersion_kx}), is a simple and often-used model to describe the dynamics of such spin qubits.
In this section, we establish the range of validity of this linear-in-$k_x$ approximation.

\begin{figure}[t]
    \centering
    \includegraphics[width=\columnwidth]{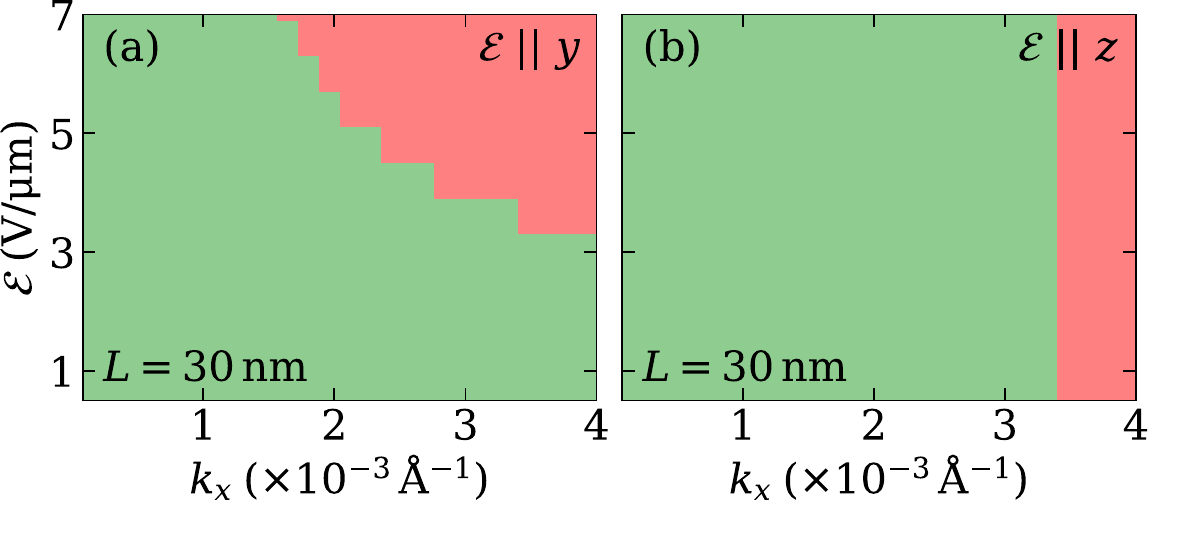}
    \caption{
Validity of the linear-in-$k_x$ (perturbative) description of the Rashba splitting for a square nanowire of side length $L = 30~\mathrm{nm}$. The spin splitting from the effective term $H_{10}k_x$ is compared with that from exact numerical diagonalization. Green regions of the $(k_x,\mathcal{E})$ plane mark where the relative deviation from the exact result is below $5\%$, i.e., where the perturbative description is accepted; red regions mark where it exceeds $5\%$. Panels (a) and (b) correspond to electric fields applied along $y$ and $z$, respectively.
    }
    \label{fig:validity_zone_linear_k}
\end{figure}

To describe an electrically defined quantum dot in a hexagonal germanium nanowire, we assume that multiple bottom gates form a parabolic electrostatic confinement potential along the nanowire axis. We estimate the $k_x$ momentum range of the ground-state orbital of such a dot here.
The relation between the orbital level spacing $\hbar\omega_{\mathrm{orb}}$ and the oscillator length $l_\mathrm{dot}$ reads:
\begin{align}
    l_{\mathrm{dot}}
    =
    \sqrt{\frac{\hbar}{|m_x^*|\omega_{\mathrm{orb}}}}.
\end{align}

For spin-qubit operation, the orbital spacing should be sufficiently large compared to the qubit splitting~\cite{kolok2025acoustic}, which is at most a few tens of GHz.
Taking the frequency, $\omega_{\mathrm{orb}}/2\pi=100~\mathrm{GHz}$ as a reasonable lower bound on the orbital level spacing, and the longitudinal effective mass $|m_x^*|=0.0737\,m_\mathrm{e}$ for a finite cross-section of $L=30~\mathrm{nm}$ (see Appendix~\ref{app:effmass} for how the longitudinal effective mass varies with cross-section) provide an upper bound
\begin{align}
    l_{\mathrm{dot}}=50~\mathrm{nm}
\end{align}
on the oscillator length and hence the spatial extension of the dot along the nanowire axis.
The standard deviation of the ground-state momentum distribution is then
\begin{align}
    \Delta k_x
    =
    \frac{1}{\sqrt{2}\,l_{\mathrm{dot}}}
    =
    1.41\times10^{-3}~\text{\AA}^{-1}.
    \label{eq:dot_momentum_width}
\end{align}
Thus, $k_x\sim 10^{-3}~\text{\AA}^{-1}$ is the characteristic momentum scale for a dot with an orbital spacing of approximately
$100~\mathrm{GHz}$. 
Hence, the linear-in-$k_x$ Rashba model provides an accurate description of spin-orbit physics for such a quantum dot as long as the electric-field-induced spin splitting is well approximated by a linear function of $k_x$ in the range $|k_x| \lesssim 10^{-3}~\text{\AA}^{-1}$.

The green regions in Figs.~\ref{fig:validity_zone_linear_k}(a) and \ref{fig:validity_zone_linear_k}(b) indicate the part of the $(k_x,\mathcal{E})$ plane where the linear-in-$k_x$ expression for the Rashba splitting matches the full numerical result within a relative error of $5\%$, whereas in the red regions the deviation exceeds $5\%$. 
The momentum range $|k_x| \lesssim 10^{-3}~\text{\AA}^{-1}$ relevant for a quantum dot resides within these green regions, validating the usage of the linear-in-$k_x$ Rashba model for quantum-dot spin qubits in these nanowires.

\subsection{Electric-field and cross-section size dependence of the Rashba coefficient and the spin-orbit length}
\label{subsec:rashba_size_dependence}

So far, we have studied how the Rashba coefficient is controlled by an electric field. Here, we consider the role of the cross-section size, since the nanowires can be grown with different side lengths.

\begin{figure}[t]
    \centering
    \includegraphics[width=\linewidth]{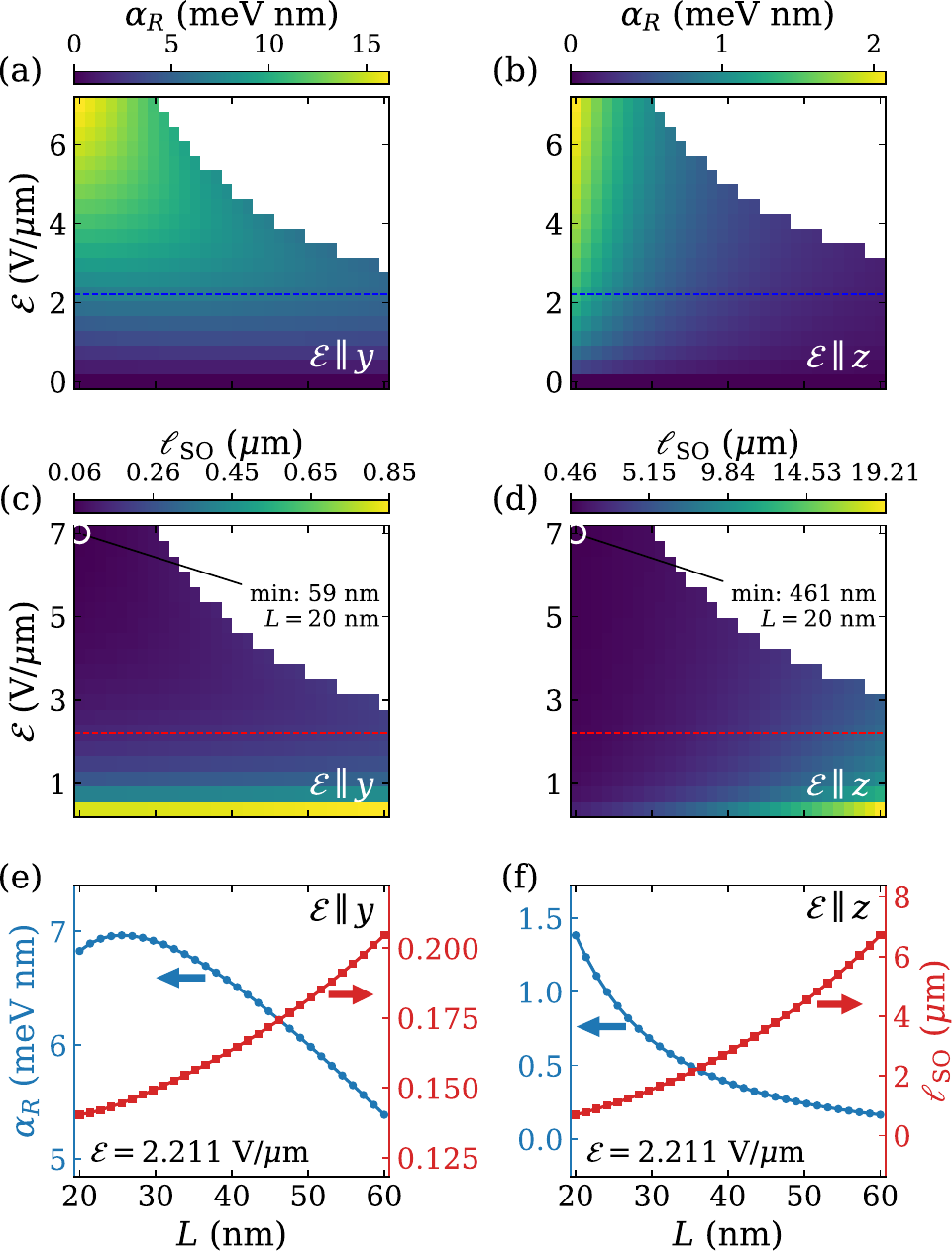}
   \caption{Electric-field and size dependence of the Rashba coefficient and spin-orbit length in hexagonal germanium nanowires. (a),(b)~Color maps of $\alpha_R$ versus cross-section side length $L$ and electric-field strength $\mathcal{E}$ for $\mathcal{E}\parallel y$ and $\mathcal{E}\parallel z$, respectively. (c),(d)~Corresponding color maps of the spin-orbit length $\ell_{\mathrm{SO}}$. The open white circles mark the minimum spin-orbit length, $\ell_{\mathrm{SO}}^{\min}\approx 59~\mathrm{nm}$ for $\mathcal{E}\parallel y$ and $\ell_{\mathrm{SO}}^{\min}\approx 461~\mathrm{nm}$ for $\mathcal{E}\parallel z$, both reached at $L=20~\mathrm{nm}$ and $\mathcal{E}=7~\mathrm{V/\mu m}$. In (a)--(d), white regions mark parameters for which the fundamental gap $E_g$ is smaller than its value at $\mathcal{E}=7~\mathrm{V/\mu m}$ and $L=30~\mathrm{nm}$. (e),(f)~Size dependence of $\alpha_R$ (blue circles, left axis) and $\ell_{\mathrm{SO}}$ (red squares, right axis) at fixed $\mathcal{E}=2.21~\mathrm{V/\mu m}$, along the dashed lines in (a)--(d).}
    \label{fig:rashba_lso}
\end{figure}

Besides the Rashba coefficient, the spin-orbit length $\ell_\mathrm{SO}$ is  often considered as an alternative characteristic of the strength of the Rashba effect.
The spin-orbit length is the length scale on which a moving spin suffers a full rotation of angle $2 \pi$; a stronger Rashba effect implies a shorter spin-orbit length.
This length scale is calculated from the Rashba coefficient $\alpha_R$ and the longitudinal effective mass $m_x$ for the given cross-sectional nanowire:
\begin{align}
    \ell_{\mathrm{SO}}
    =
    \frac{\hbar^2}{|m_x|\alpha_R}.
\end{align}
Figure~\ref{fig:rashba_lso} shows how $\alpha_R$ and the corresponding $\ell_{\mathrm{so}}$ depend on the electric-field strength $\mathcal{E}$ and the cross-section size $L$, for $20~\mathrm{nm}\le L\le 60~\mathrm{nm}$ and fields up to $7~\mathrm{V/\mu m}$. Figures~\ref{fig:rashba_lso}(a) and~\ref{fig:rashba_lso}(b) show $\alpha_R$ for $\mathcal{E}\parallel y$ and $\mathcal{E}\parallel z$, respectively, and Figs.~\ref{fig:rashba_lso}(c) and~\ref{fig:rashba_lso}(d) show the corresponding $\ell_{\mathrm{so}}$.

Because the quadratic Stark shift grows with both the field strength and the cross-section size, the fundamental gap $E_g$ shrinks toward the upper-right corner of each map. To stay in the same gapped regime as in the preceding subsections, we take the gap at $\mathcal{E}=7~\mathrm{V/\mu m}$ and $L=30~\mathrm{nm}$ as a reference. This gives the threshold value, $E_g^{y}=0.21~\mathrm{eV}$ for $\mathcal{E}\parallel y$ and $E_g^{z}=0.167~\mathrm{eV}$ for $\mathcal{E}\parallel z$. Parameters for which $E_g$ falls below the reference value are excluded and shown in white.

For both field orientations, $\alpha_R$ increases with increasing field strength over most of the parameter range, though not strictly monotonically. Since $\ell_{\mathrm{SO}}$ is inversely proportional to $\alpha_R$, it shows the opposite trend and decreases with increasing field. At strong fields, $\alpha_R$ is largest in the narrowest wires. At $L=20~\mathrm{nm}$ and $\mathcal{E}=7~\mathrm{V/\mu m}$, it reaches $\approx 16~\mathrm{meV\,nm}$ for $\mathcal{E}\parallel y$, which corresponds to a minimum spin-orbit length of $\ell_{\mathrm{SO}}\approx 59~\mathrm{nm}$. For $\mathcal{E}\parallel z$, it reaches only $\approx 2~\mathrm{meV\,nm}$, which corresponds to $\ell_{\mathrm{SO}}\approx 461~\mathrm{nm}$.

Figures~\ref{fig:rashba_lso}(e) and~\ref{fig:rashba_lso}(f) show cuts through the maps at the fixed field $\mathcal{E}=2.21~\mathrm{V/\mu m}$ (dashed lines in Figs.~\ref{fig:rashba_lso}(a)--\ref{fig:rashba_lso}(d)). At this field, the full size range lies within the allowed region. In these panels, $\alpha_R$ is shown by blue circles (left axis) and $\ell_{\mathrm{SO}}$ by red squares (right axis).

The two orientations depend on size in different ways. For $\mathcal{E}\parallel y$, $\alpha_R$ first rises slightly with increasing $L$, reaches a broad maximum of $\approx 7.0~\mathrm{meV\,nm}$ near $L\simeq 25~\mathrm{nm}$, and then decreases to $\approx 5.4~\mathrm{meV\,nm}$ at $L=60~\mathrm{nm}$. Across the whole range it varies by less than $25\%$. Over the same range, $\ell_{\mathrm{SO}}$ increases from $\approx 0.14~\mathrm{\mu m}$ to $\approx 0.21~\mathrm{\mu m}$. For $\mathcal{E}\parallel z$, in contrast, $\alpha_R$ decreases monotonically with increasing $L$ by a factor of about eight, from $\approx 1.4~\mathrm{meV\,nm}$ at $L=20~\mathrm{nm}$ to $\approx 0.17~\mathrm{meV\,nm}$ at $L=60~\mathrm{nm}$. Correspondingly, $\ell_{\mathrm{SO}}$ increases from $\approx 0.8~\mathrm{\mu m}$ to $\approx 6.8~\mathrm{\mu m}$. In both orientations, $\ell_{\mathrm{SO}}$ therefore grows with the wire size. For $\mathcal{E}\parallel y$, however, it is shorter by a factor of about $6$ at $L=20~\mathrm{nm}$ and about $30$ at $L=60~\mathrm{nm}$. Over the full maps in Figs.~\ref{fig:rashba_lso}(c) and~(d), the minimum spin-orbit length is $\ell_{\mathrm{SO}}\approx 59~\mathrm{nm}$ for $\mathcal{E}\parallel y$ and $\ell_{\mathrm{SO}}\approx 461~\mathrm{nm}$ for $\mathcal{E}\parallel z$, both reached at $L=20~\mathrm{nm}$ and the largest field considered.

These results identify field orientation, field strength, and cross-section size as complementary parameters for tuning the Rashba interaction while keeping the fundamental gap open. A field perpendicular to the $c$ axis gives the stronger and more size-robust response, and as the field strength increases, the size at which $\alpha_R$ is maximal shifts toward smaller $L$. A field parallel to the $c$ axis gives a much weaker response, which is further suppressed as the cross section grows. Narrow wires are therefore preferable when the gate field is mainly along $c$. We note that Fig.~\ref{fig:rashba_lso} considers fields strictly along $y$ or $z$. At strong fields, tilting the field slightly from the $y$ axis toward the $c$ axis enhances $\alpha_R$ further [Fig.~\ref{fig:Fig-5}(c)].

\section{Microscopic origin of the Rashba coupling}
\label{symmetry_analysis}

The finite-field treatment in Sec.~\ref{sec:finite_field} captures the nonlinear electric-field dependence of the Rashba coupling but does not identify the confined states responsible for it. To resolve its microscopic origin, we now expand perturbatively in both the transverse electric field and the longitudinal momentum $k_x$, focusing on the leading weak-field contribution.

We take $H_0(k_x=0)$ as the unperturbed Hamiltonian and treat the electric-dipole term $H_{\mathcal E_i}$ [Eq.~\eqref{eq:electric_field_matrix_form}], the linear-in-$k_x$ term $k_xV_1$ [Eq.~\eqref{V1}], and the quadratic term $k_x^2V_2$ as perturbations, where $i\in\{y,z\}$. Using a Schrieffer--Wolff transformation implemented in \textsc{Pymablock}~\cite{10.21468/SciPostPhysCodeb.50-r2.1}, we project the Hamiltonian onto the Kramers doublet of the top valence subband. Following the notation introduced in Sec.~\ref{subsec:finite_field_rashba}, we label the effective coefficients by $H_{abc}$, where the indices count insertions of the electric-field perturbation, $k_xV_1$, and $k_x^2V_2$, respectively.

Retaining terms first order in both $\mathcal E_i$ and $k_x$, the leading pseudospin-dependent contribution is
\begin{align}
    H_{R,\mathrm{eff}}^{(i)}
    = H_{110}^{(i)}\,\mathcal E_i k_x,
    \label{eq:perturbative_rashba_hamiltonian}
\end{align}
which arises from virtual inter-subband processes involving one electric-dipole insertion and one insertion of $V_1$. The resulting Rashba coefficient is linear in the electric field in the weak-field limit. We therefore define the field-normalized coefficient as
\begin{align}
    \alpha_R^{\prime(i)}
    \equiv
    \left.
    \frac{\partial\alpha_R^{(i)}}{\partial\mathcal E_i}
    \right|_{\mathcal E_i=0},
    \qquad i\in\{y,z\}.
    \label{eq:alpha_prime}
\end{align}
This coefficient corresponds to the initial slope of the finite-field results in Figs.~\ref{fig:Fig-5}(a,b) and quantifies the electrical tunability of the Rashba interaction. It also provides a convenient basis for comparison with other semiconductor nanowire platforms (Sec.~\ref{sec:rashba_comparison}).

The pronounced anisotropy of the Rashba response raises a further question: which confined subbands mediate the effective spin-orbit coupling, and what is their dominant bulk-band character? To address this question, we resolve the mixed-order coefficient $H_{110}^{(i)}$ into contributions from individual intermediate states. We denote the two Kramers partners of the top valence subband at $k_x=0$ by $|v_\uparrow\rangle$ and $|v_\downarrow\rangle$, with energy $E_{\mathrm{vb}_0}$. These states span the low-energy subspace $A$, while its orthogonal complement is denoted by $B$.

Then the effective Hamiltonian in the top-valence-subband subspace can be written as
\begin{align}
    H_R^{(i)}
    =
    -e\mathcal{E}_i k_x
    \sum_{n\in B}
    \begin{pmatrix}
        \mathcal{A}_{\uparrow\uparrow}^{(i,n)}
        &
        \mathcal{A}_{\uparrow\downarrow}^{(i,n)}
        \\
        \mathcal{A}_{\downarrow\uparrow}^{(i,n)}
        &
        \mathcal{A}_{\downarrow\downarrow}^{(i,n)}
    \end{pmatrix}.
    \label{eq:HR_SW_sum}
\end{align}
The contribution from an intermediate subband $|n\rangle$ is
\begin{equation}
    \mathcal{A}_{\sigma\sigma'}^{(i,n)}
    =
    \frac{\langle v_\sigma|r_i|n\rangle
    \langle n|V_1|v_{\sigma'}\rangle
    +\langle v_\sigma|V_1|n\rangle
    \langle n|r_i|v_{\sigma'}\rangle}{E_{\mathrm{vb}_0} - E_n}
   \label{eq:SW_individual_contribution}
\end{equation}
where $E_{\mathrm{vb}_0}$ is the energy of the top confined valence subband at $k_x=0$ and $E_n$ is the energy of the intermediate subband $|n\rangle$.

Eq.~\eqref{eq:SW_individual_contribution} makes explicit that a subband contributes to the Rashba splitting only if it is connected to the top valence subband by both the electric-dipole operator $r_i$ and the $k_x$-dependent operator $V_1$. The relative contribution of each intermediate state is therefore controlled by three factors: the electric-dipole matrix element, the $k_x$-dependent matrix element, and the energy denominator.

\subsection{Selection rules}
\label{sec:selection_rules}

The symmetry origin of the allowed matrix elements can be understood by classifying the confined states at $k_x=0$. In the $\mathbf{k}\cdot\mathbf{p}$ limit, the wire Hamiltonian is insensitive to the fractional translations associated with screw rotations.
The relevant symmetry operations of the wire geometry are therefore generated by inversion $\mathcal{I}$, two-fold rotations $C_{2\alpha}$ about the coordinate axes $(\alpha=x,y,z)$, and time reversal $\mathcal{T}$.
We use the symmetries of the wire represented according to the ten-band basis listed in Appendix~\ref{app:hamiltonian}.
The two-fold rotation around the $z$ axis and inversion operators are represented on the internal degrees of freedom by~\cite{Chunag96,pulcu2024multiband,Claudia2019accurate,Varjas_2018}
\begin{align}
U_{C_{2z}}
&=
- i\,\mathrm{diag}
(+1,+1,-1,-1,+1,-1,-1,+1,+1,-1),
\\
U_\mathcal{I}
&=
\phantom{- i\,}\mathrm{diag}
(-1,-1,+1,+1,+1,-1,-1,+1,+1,+1).
\end{align}
The real-space coordinates and momenta transform in the usual way. For example, $C_{2z}$ sends
\begin{align}
(x,y,z)&\rightarrow(-x,-y,z),
\\
(k_x,k_y,k_z)&\rightarrow(-k_x,-k_y,k_z),
\end{align}
while inversion sends all coordinates and momenta to their negatives.

Since the full symmetry group is non-Abelian, it is convenient to classify the eigenstates using a maximal commuting subgroup generated by $C_{2z}$ and $\mathcal{I}$. At $k_x=0$, the eigenstates can therefore be labeled by two quantum numbers,
\begin{align}
C_{2z}|\psi_n\rangle
&=
\lambda_{C_{2z}}^{(n)}|\psi_n\rangle,
&
\mathcal{I}|\psi_n\rangle
&=
\lambda_\mathcal{I}^{(n)}|\psi_n\rangle,
\label{eq:symmetry_labels}
\end{align}
with
\begin{align}
\lambda_{C_{2z}}^{(n)}=\pm i,
\qquad
\lambda_\mathcal{I}^{(n)}=\pm 1 .
\end{align}
These quantum numbers refer to the full ten-component confined-state wave function; the total symmetry label contains both the symmetry of the internal ten-band orbital component and the symmetry of the real-space envelope. Schematically, for $g=C_{2z},\mathcal{I}$,
\begin{align}
    \lambda_g^{\mathrm{tot}}
    =
    \lambda_g^{\mathrm{int}}
    \lambda_g^{\mathrm{env}} .
    \label{eq:total_sym_eigenvalue}
\end{align}
Therefore, two confined subbands with the same dominant bulk-band character do not necessarily have the same total symmetry quantum numbers if their envelope functions have different parity. This distinction is important below: the selection rules are determined by the total quantum numbers of the confined states, not by their internal orbital character alone.

For example, the top valence subband, denoted here as vb$_0$, has a symmetric real-space envelope, and its dominant internal orbital component, originating from the top bulk valence band, is also even under inversion. Hence, the total inversion parity of the top vb$_0$ doublet is
\begin{align}
\lambda_\mathcal{I}^{(v_{\uparrow,\downarrow})}=+1 .
\end{align}
We denote the two Kramers partners by $|\widetilde{v}_\uparrow\rangle$ and
$|\widetilde{v}_\downarrow\rangle$, with opposite $C_{2z}$ eigenvalues,
\begin{align}
\lambda_{C_{2z}}^{(v_\uparrow)}=+i,
\qquad
\lambda_{C_{2z}}^{(v_\downarrow)}=-i .
\end{align}
In order to fix the relative phase of the states within the Kramers pair, we use the mirror $M_x$ with normal along the $x$ axis, represented by
\begin{equation}
    U_{M_x} = - i\,\sigma_x \otimes \mathbb{I}_5.
\end{equation}
For the highest valence subband, the envelope functions are fully symmetric, so we may ignore the real-space part of this operator.
The relative phase is fixed by setting
\begin{align}
    \ket{v_\downarrow}
    =
    -i \left(\langle\widetilde{v}_\uparrow|U_{M_x}|\widetilde{v}_\downarrow\rangle\right)^*
    \ket{\widetilde{v}_\downarrow}, \qquad \ket{v_\uparrow} = \ket{\widetilde{v}_\uparrow}.
\end{align}
With this convention,
$i \langle v_\uparrow|U_{M_x}|v_\downarrow\rangle$ is real and positive, and
the $M_x$ mirror operator restricted to the Kramers subspace
takes the form $\Psi^\dagger U_{C_{M_x}}\Psi= -i \sigma_x$, with $\Psi=\bigl(\ket{v_\uparrow},\ket{v_\downarrow}\bigr)$.

The Rashba splitting is generated by a second-order virtual process involving both the transverse electric-dipole perturbation and the part of the Hamiltonian linear in $k_x$. We write the linear-in-$k_x$ term as 
\begin{align}
H^{(1)}(k_x)=V_1k_x .
\end{align}
The full product $V_1k_x$ is symmetry-allowed. However, under both inversion and $C_{2z}$, the momentum $k_x$ changes sign. Therefore, the matrix coefficient $V_1$ itself must also be odd under both operations, so that the product $V_1k_x$ remains invariant. The symmetry characters of the operators relevant to the second-order Rashba process are summarized in
Table~\ref{tab:operator_symmetry}.

\begin{table}[t]
    \caption{
    Symmetry characters of the operators entering the second-order Rashba process. Here $V_1$ denotes the coefficient of the linear-in-$k_x$ contribution $V_1k_x$.
    }
    \label{tab:operator_symmetry}

    \centering
    \renewcommand{\arraystretch}{1.15}

    \begin{tabular*}{\columnwidth}{
        @{\extracolsep{\fill}}
        l
        c
        c
        @{}
    }
        \toprule
        \toprule

        Operator
        &
        $C_{2z}$
        &
        $\mathcal{I}$
        \\

        \midrule

        $V_1$ & $-$ & $-$
        \\

        $y$ & $-$ & $-$
        \\

        $z$ & $+$ & $-$
        \\

        \bottomrule
        \bottomrule
    \end{tabular*}
\end{table}

For a transverse electric field along $y$, the perturbation is $e\mathcal{E}_y y$. From Table~\ref{tab:operator_symmetry}, both $y$ and $V_1$ are odd under $C_{2z}$ and $\mathcal{I}$:
\begin{align}
y:\;(-,-),
\qquad
V_1:\;(-,-).
\end{align}
Starting from the top valence subband state $|v_{\uparrow,\downarrow}\rangle$, the electric-dipole operator therefore couples only to intermediate states whose total symmetry quantum numbers satisfy
\begin{align}
\lambda_{C_{2z}}^{(n)}=\mp i,
\qquad
\lambda_\mathcal{I}^{(n)}=-1 .
\end{align}
The same symmetry sector is also selected by $V_1$. Consequently, the allowed second-order paths have the form
\begin{align}
|v_\uparrow; +i,+1 \rangle
&\overset{\;e\mathcal{E}_y y\;}{\longleftrightarrow}
|n; -i,-1 \rangle
\overset{\;V_1\;}{\longleftrightarrow}
|v_\uparrow; +i,+1 \rangle ,
\\
|v_\downarrow; -i,+1 \rangle
&\overset{\;e\mathcal{E}_y y\;}{\longleftrightarrow}
|n; +i,-1 \rangle
\overset{\;V_1\;}{\longleftrightarrow}
|v_\downarrow; -i,+1 \rangle .
\end{align}
Thus the $y$-field-induced Rashba term is diagonal in the $C_{2z}$-labeled vb$_0$ doublet and has the form
\begin{align}
H_R^{(y)}
\propto k_x\mathcal{E}_y \sigma_z .
\end{align}

This selection rule can be satisfied not only by subbands with different internal orbital character, such as conduction-band-like states, but also by VB$_1$-like excited subbands whose envelope parity changes the total $(C_{2z},\mathcal{I})$ quantum numbers. For instance, a VB$_1$-like state with an envelope odd in $y$ has the correct total symmetry to be connected to the top valence subband state by the $y$-dipole operator.

For a transverse electric field along $z$, the perturbation is $e\mathcal{E}_z z$. In this case $z$ is even under $C_{2z}$ but odd under inversion, while $V_1$ remains odd under both:
\begin{align}
z:\;(+,-),
\qquad
V_1:\;(-,-).
\end{align}
The electric-dipole operator $z$ therefore connects $|v_{\uparrow,\downarrow}\rangle$ to intermediate states whose total symmetry quantum numbers satisfy
\begin{align}
\lambda_{C_{2z}}^{(n)}=\pm i,
\qquad
\lambda_\mathcal{I}^{(n)}=-1 .
\end{align}
The subsequent action of $V_1$ flips the $C_{2z}$ eigenvalue and connects this intermediate state to the opposite member of the vb$_0$ doublet. The allowed virtual processes are therefore
\begin{align}
|v_\uparrow; +i,+1 \rangle
&\overset{\;e\mathcal{E}_z z\;}{\longleftrightarrow}
|n; +i,-1 \rangle
\overset{\;V_1\;}{\longleftrightarrow}
|v_\downarrow; -i,+1 \rangle ,
\\
|v_\downarrow; -i,+1 \rangle
&\overset{\;e\mathcal{E}_z z\;}{\longleftrightarrow}
|n; -i,-1 \rangle
\overset{\;V_1\;}{\longleftrightarrow}
|v_\uparrow; +i,+1 \rangle .
\end{align}
Hence, the $z$-field-induced Rashba term is off-diagonal in the vb$_0$ doublet and
takes the form
\begin{align}
H_R^{(z)}
\propto
 k_x\mathcal{E}_z \sigma_y .
\end{align}
This naturally explains why a VB$_1$-like subband with lobes along the $z$ direction can strongly contribute to the $z$-field-induced Rashba splitting: although its internal orbital character is similar to that of the top VB$_1$ state, its $z$-odd envelope flips the total inversion parity while preserving the $C_{2z}$ label, exactly as required by the $z$-dipole selection rule.

\begin{figure}[t]
    \centering
    \includegraphics[width=\columnwidth]{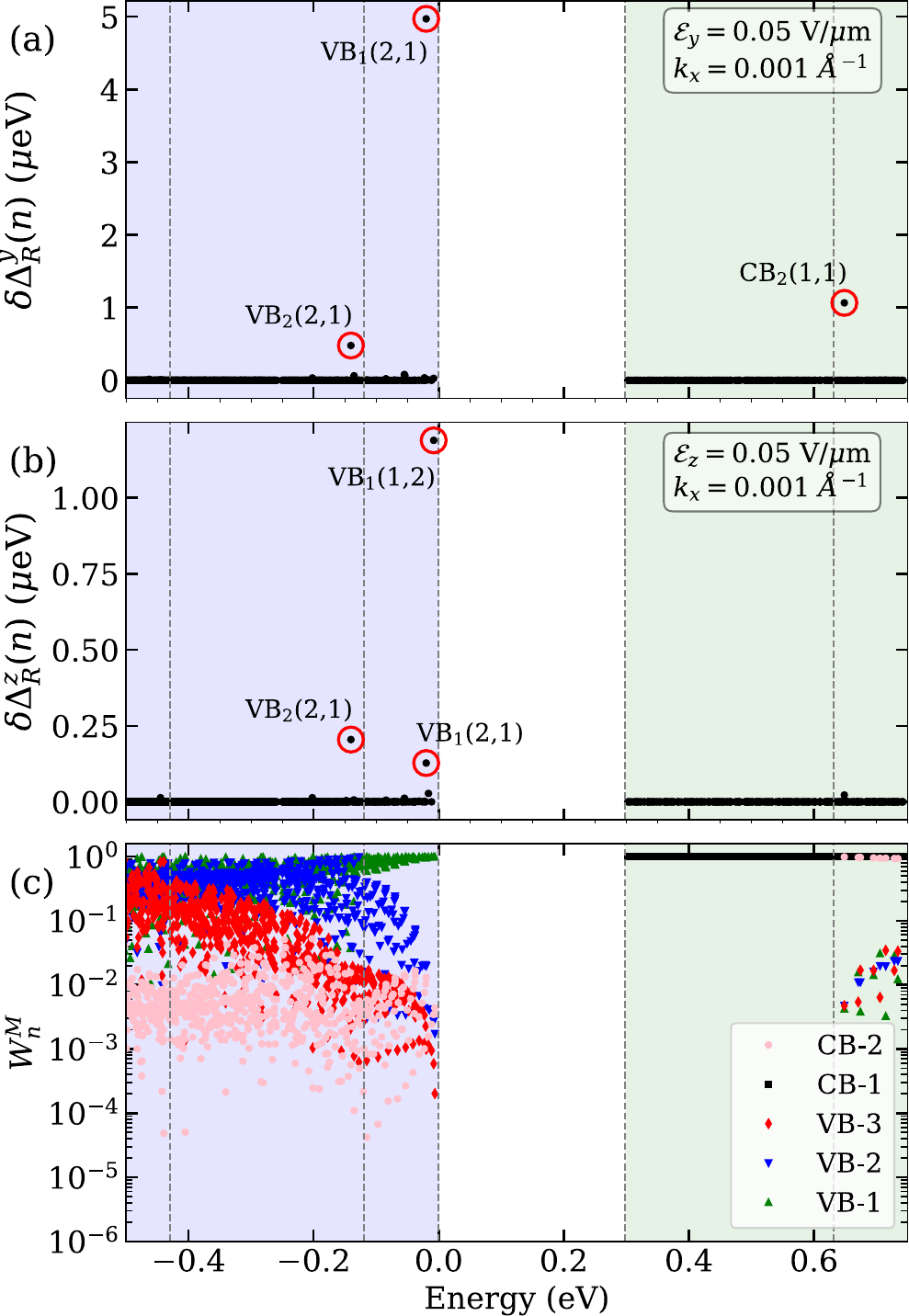}
    \caption{
Subband-resolved origin of the Rashba splitting for a nanowire with cross-section side length $L=30~\mathrm{nm}$. (a,b) Kramers-pair-resolved complementary-subspace contributions $\delta\Delta_R^{(i)}(n)$ for electric fields along (a) $y$ and (b) $z$, calculated at $k_x=0.001~\mathrm{\AA}^{-1}$ and $\mathcal{E}_y=\mathcal{E}_z=0.05$ V/$\mu$m for the respective panels. Each point corresponds to an intermediate Kramers pair and represents the sum of the contributions from its two constituent states. Red circles highlight the dominant intermediate Kramers pairs. The blue and green-shaded regions indicate the valence and conduction band subspaces, respectively, while the white region denotes the bulk band gap. (c) Bulk band character weights $W_n^M$ of the confined subbands, obtained by projecting them onto the five relevant bulk-band components at $k_x=0$, revealing the dominant band characters associated with the Rashba response.
}
    \label{fig:overlap}
\end{figure}

\subsection{Dominant intermediate subbands}

\begin{table}[t]
    \caption{
    Symmetry eigenvalues of the dominant subband levels are shown in Fig.~\ref{fig:level}. The first column gives the dominant character of each confined nanowire subband. The notation $\mathrm{VB}_\mu(n_y,n_z)$ or
    $\mathrm{CB}_\mu(n_y,n_z)$ indicates that the confined state is mainly derived from the bulk band $\mathrm{VB}_\mu$ or $\mathrm{CB}_\mu$ and has an approximate transverse envelope character $\sin[n_y\pi(y+L/2)/L]\sin[n_z\pi(z+L/2)/L]$. Thus,
    $\mathrm{VB}_1(1,1)$ denotes the top confined valence subband, while $\mathrm{VB}_1(1,2)$ and $\mathrm{VB}_1(2,1)$ denote excited confined valence-sector states with dominant $z$- and $y$-odd envelope character, respectively. Each level forms a Kramers pair, and the two columns denote the two $C_{2z}$-labeled branches, written using the shorthand labels $\ket{\uparrow}$ and $\ket{\downarrow}$. These labels should not be interpreted as pure spin eigenstates in the presence of spin-orbit coupling. The entries give the corresponding eigenvalues of $C_{2z}$ and inversion $\mathcal{I}$.
    }
    \label{tab:level_symmetry}

    \centering
    \renewcommand{\arraystretch}{1.15}

    \begin{tabular*}{\columnwidth}{
        @{\extracolsep{\fill}}
        l
        c
        c
        @{}
    }
        \toprule
        \toprule

        Dominant character
        &
        $\ket{\uparrow} \mathrm{branch} \;(C_{2z},\mathcal{I})$
        &
        $\ket{\downarrow} \mathrm{branch} \;(C_{2z},\mathcal{I})$
        \\

        \midrule

        $\mathrm{VB}_1(1,1)$
        &
        $(+i,+1)$
        &
        $(-i,+1)$
        \\

        $\mathrm{VB}_1(1,2)$
        &
        $(+i,-1)$
        &
        $(-i,-1)$
        \\

        $\mathrm{VB}_1(2,1)$
        &
        $(-i,-1)$
        &
        $(+i,-1)$
        \\

        $\mathrm{VB}_2(2,1)$
        &
        $(-i,-1)$
        &
        $(+i,-1)$
        \\

        $\mathrm{CB}_2(1,1)$
        &
        $(-i,-1)$
        &
        $(+i,-1)$
        \\

        \bottomrule
        \bottomrule
    \end{tabular*}
\end{table}

To identify the microscopic origin of the Rashba coupling, we decompose the Schrieffer--Wolff correction into contributions from individual intermediate Kramers pairs in the complementary subspace. We label each pair by $n$. The combined
contribution of the pair to the Rashba splitting of the top valence subband is denoted by $\delta\Delta_R^{(i)}(n)$, where $i=y,z$ specifies the direction of the applied electric field. Operationally, $\delta\Delta_R^{(i)}(n)$ is obtained from Eq.~\eqref{eq:HR_SW_sum} by retaining both the Kramers pair states in the Schrieffer-Wolff sum. Thus, the reported quantity represents the sum of the contributions from the two members of the intermediate Kramers pair, rather than the contribution of an individual state.

Figs.~\ref{fig:overlap}(a) and \ref{fig:overlap}(b) show $\delta\Delta_R^{(i)}(n)$ as a function of the intermediate-subband energy for electric fields applied along $y$ and $z$, respectively, with $\mathcal{E}=0.05~\mathrm{V/\mu m}$. Each plotted point corresponds to one intermediate Kramers pair. In both field directions, only a small number of pairs give sizable contributions compared with the rest of the spectrum. These dominant Kramers pairs are highlighted by red circles.

For $\mathcal{E}\parallel y$, the dominant intermediate Kramers pairs include two valence-sector pairs with energies at
\begin{align}
    E_n \simeq -0.02002~\mathrm{eV},
    \qquad
    E_n \simeq -0.1404~\mathrm{eV},
\end{align}
and one conduction-sector pair at
\begin{align}
    E_n \simeq 0.64863~\mathrm{eV}.
\end{align}
For $\mathcal{E}\parallel z$, the dominant intermediate Kramers pairs lie entirely in the valence sector. The largest contribution comes from the pair closest in energy to the top valence subband, at \begin{align}
    E_n \simeq -0.00805~\mathrm{eV}.
\end{align}
The other two important valence-sector pairs occur at approximately the same energies as those identified for $\mathcal{E}\parallel y$. The corresponding real-space probability densities and symmetry assignments are discussed in
Appendix~\ref{app:dominant_subband}.

The bulk-band character of the dominant subbands is shown in Fig.~\ref{fig:overlap}(c). We quantify this character by projecting each nanowire eigenstate $\psi_n$ onto the bulk-band eigenstates at $\Gamma$,
\begin{align}
    W_n^M
    =
    \sum_{y,z}
    \sum_{s \in \{\uparrow, \downarrow\}}
    \left|
    \sum_{a=1}^{10}
    \left[u_{a}^{(M,s)}\right]^*
    \psi_{n,a}(y,z)
    \right|^2 ,
    \label{eq:bulk_weight_main}
\end{align}
where $M$ labels a bulk-band manifold and $a$ denotes the internal ten-band degree of freedom (See Appendix~\ref{app:overlap} for the full projection procedure). The dominant conduction-sector Kramers pair has mainly CB$_2$ character, whereas the dominant valence-sector pairs originate from different valence-band components. Therefore, the dominant Rashba
response is controlled not only by the energy separation from the top vb$_0$ subband, but also by the orbital composition and symmetry of the intermediate Kramers pairs.

\begin{figure}[t]
    \centering
    \includegraphics[width=\columnwidth]{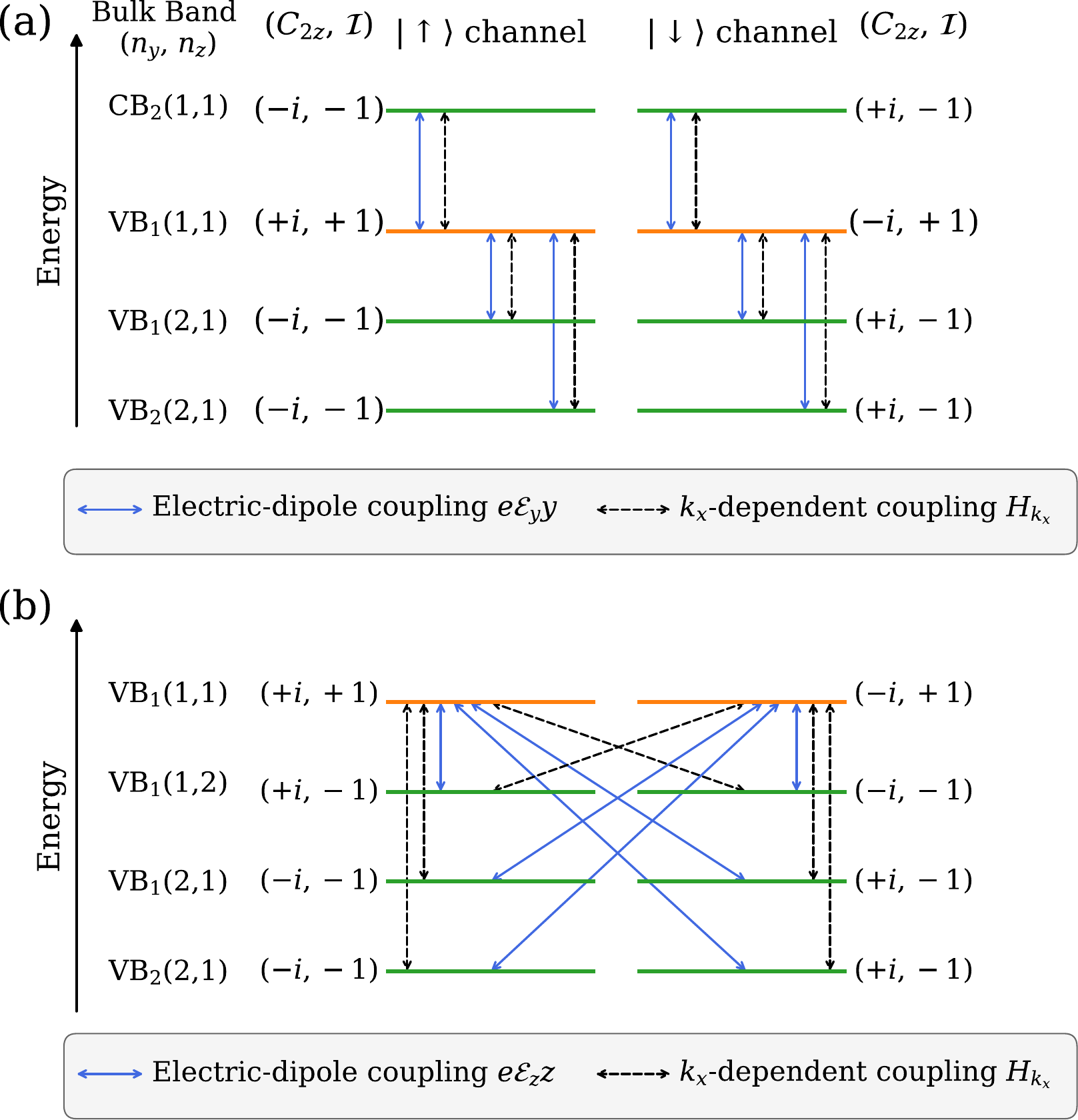}
    \caption{
    Symmetry-allowed virtual processes contributing to the Rashba splitting of the top vb$_0$ subband. Orange lines denote the two Kramers partners of the top vb$_0$ subband, while the green lines represent the Kramers partners of the dominant intermediate subbands. Blue double arrows indicate allowed electric-dipole matrix elements, whereas black dashed double arrows indicate allowed matrix elements of the $k_x$-dependent operator $V_1$. (a) For $\mathcal{E}\parallel y$, both $e\mathcal{E}_y y$ and $V_1$ flip the $(C_{2z},\mathcal{I})$ labels. The second-order process therefore returns to the same vb$_0$ Kramers partner and produces a diagonal Rashba term. (b) For $\mathcal{E}\parallel z$, the $z$-dipole operator preserves $C_{2z}$ but flips inversion, whereas $V_1$ flips both labels. Consequently, the process connects opposite vb$_0$ Kramers partners and produces an off-diagonal Rashba term.
    }
    \label{fig:level}
\end{figure}

The symmetry quantum numbers of the dominant subbands are summarized in Table~\ref{tab:level_symmetry} (See Appendix \ref{app:dominant_subband} for the detailed derivation). Using these quantum numbers, we construct the level diagrams shown in Fig.~\ref{fig:level}. These diagrams indicate which members of the intermediate Kramers pairs are connected to the top vb$_0$ doublet through the electric-dipole operator and the $k_x$-dependent operator $V_1$.

For $\mathcal{E}\parallel y$, shown in Fig.~\ref{fig:level}(a), both $e\mathcal{E}_y y$ and $V_1$ are odd under $C_{2z}$ and inversion $\mathcal{I}$. Therefore, both operators flip the symmetry labels $(C_{2z},\mathcal{I})$. The two-step virtual process connects a given vb$_0$ Kramers partner to an allowed member of an intermediate Kramers pair and then
back to the same vb$_0$ partner. The resulting Rashba term is therefore diagonal in the $C_{2z}$-labeled vb$_0$ doublet.

For $\mathcal{E}\parallel z$, shown in Fig.~\ref{fig:level}(b), the $z$-dipole operator is even under $C_{2z}$ but odd under inversion, whereas $V_1$ is odd under both symmetries. Hence, the $z$-dipole operator first connects a vb$_0$
Kramers partner to an intermediate state with the same $C_{2z}$ eigenvalue but opposite inversion parity. The operator $V_1$ then connects this intermediate state to the opposite vb$_0$ Kramers partner. The resulting Rashba
term is therefore off-diagonal in the vb$_0$ doublet.

Together, Figs.~\ref{fig:overlap} and \ref{fig:level} show that the Rashba splitting is controlled by both symmetry and microscopic details. The selection rules determine which virtual channels through the intermediate Kramers pairs are allowed, while the magnitude of each pair contribution is set by the corresponding electric-dipole and $V_1$ matrix elements, the
orbital composition of the intermediate states, and their energy separation from the top vb$_0$ subband.

\subsection{Relevance of interband couplings to the Rashba splitting}
\label{subsec:interband_relevance}
\begin{table}[t]
\caption{Relative change $\delta$ of the Rashba splitting $\Delta E$ [Eq.~\eqref{eq:linearenergysplitting}] when one of the four $\mathbf{k}\cdot\mathbf{p}$ parameters coupling the topmost valence band to the other bands is set to zero, for an electric field along $y$ ($\mathcal{E}_y$) and along $z$ ($\mathcal{E}_z$). Here $\delta = |\Delta E - \Delta E^{(0)}|/\Delta E^{(0)}$, where $\Delta E^{(0)}$ is the splitting with all parameters included: $\Delta E^{(0)} = 3.48\,\mu\mathrm{eV}$ ($\mathcal{E}_y$) and $0.81\,\mu\mathrm{eV}$ ($\mathcal{E}_z$). Since $\Delta E = 2\alpha_R|k_x|$ at fixed $k_x$, $\delta$ is also the relative change of $\alpha_R$. Here, $\mathcal{E} = 0.05\,\mathrm{V}/\mu\mathrm{m}$, $k_x = 0.001\,\mathrm{\AA}^{-1}$, $L = 30\,\mathrm{nm}$.}
\label{tab:rashba_params}
\begin{ruledtabular}
\begin{tabular}{lcc}
Parameter switched off & $\delta$ for $\mathcal{E}_y$ (\%) & $\delta$ for $\mathcal{E}_z$ (\%) \\
\hline
$p_\perp = 0$        & 77.29 & 97.34 \\
$A_5 = 0$      & 80.25 & 20.29 \\
$A_6 = 0$      & 1.09  & 26.62 \\
$\Delta_3 = 0$ & 33.69 & 100.00 \\
\end{tabular}
\end{ruledtabular}
\end{table}

\begin{table*}[t]
    \caption{
    Comparison of electrically induced Rashba spin-orbit coupling in semiconductor nanowire systems and other group-IV hole nanostructures. The labels (T) and (E) denote theoretical and experimental results, respectively. The field-normalized coefficient is defined as $\alpha_R'=\partial\alpha_R/\partial\mathcal{E}$ in the approximately linear-response regime. The quantities $\alpha_R^{\max}$ and $\ell_{\mathrm{SO}}^{\min}$ denote the largest Rashba coefficient and shortest spin-orbit length reported in, or inferred from, each study for the specified system within the investigated parameter range. Dashes indicate quantities not specified or not extracted for this comparison. Here, $L$ denotes the side length of a square cross section, $d$ the diameter of a circular cross section, $t_s$ is the shell thickness, $W$ the width of a hexagonal cross section, and $t\times w$ the thickness and width of a rectangular channel. The lower block lists cubic Si and Ge hole quantum-dot devices, where the definition of $\ell_{\mathrm{SO}}$ differs between experiments (see footnotes d--h), so these values are indicative only.
    }
    \label{tab:rashba_comparison}

    \small
    \renewcommand{\arraystretch}{1.20}
    \setlength{\tabcolsep}{5pt}

    \begin{tabular}{llcccc}
        \toprule
        \toprule

        Material
        &
        System
        &
        \shortstack{Characteristic\\transverse size (nm)}
        &
        \shortstack{$\alpha_R'$\\($e\,\mathrm{nm}^2$)}
        &
        \shortstack{$\alpha_R^{\max}$\\($\mathrm{meV\,nm}$)}
        &
        \shortstack{$\ell_{\mathrm{SO}}^{\min}$\\($\mathrm{nm}$)}
        \\

        \midrule

        2H-Ge (T), $\mathcal{E}\parallel y$
        & NW
        & $L=30$
        & $\sim3.48$
        & $\sim11.2$
        & $\sim92.29$
        \\

        2H-Ge (T), $\mathcal{E}\parallel z$
        & NW
        & $L=30$
        & $\sim0.81$
        & $\sim1.0$
        & $\sim1033.7$
        \\

        2H-Ge (T), $\theta \simeq 83.2^\circ$
        & NW
        & $L=30$
        & ---
        & $\sim16.44$
        & $\sim62.87$
        \\

        Ge/Si (T) \cite{kloeffel2018direct}
        & Core/shell NW
        & $d = 10, \,t_s = 1.5$
        & $\sim8.4$
        & $\sim69.7$
        & $\sim23.24$
        \\

        InSb (T)\textsuperscript{a}
        \cite{zhang2006rashba}
        & NW
        & $d=48$
        & $\sim3.6$
        & $\sim23$
        & $\sim236.6$
        \\

        InSb (E)\textsuperscript{b}
        \cite{kammhuber2017conductance}
        & NW
        & $d\simeq100$
        & ---
        & $\sim270$
        & $\sim20$
        \\

        InAs (E)\textsuperscript{c}
        \cite{liang2012strong}
        & NW
        & $d\simeq40$
        & $\sim1.17$
        & $\sim32$
        & $\sim50$
        \\

        GaAs (T)
        \cite{wojcik2018tuning}
        & NW
        & $W=87$
        & ---
        & $\sim4.6$
        & $\sim247$
        \\

        Ge/Si (E) \cite{hao2010strong}
        & Core/shell NW
        & $d = 10, \,t_s = 2$
        & ---
        & $\sim108$
        & $\sim12$
        \\

        \midrule

        Si (E)\textsuperscript{d,\dag}
        \cite{Li2015NanoLett}
        & Planar MOS DQD
        & ---
        & ---
        & ---
        & $\sim110$
        \\

        Si (E)\textsuperscript{e,*}
        \cite{Yu2023NatNano}
        & NW DQD
        & $t\times w=10\times40$
        & ---
        & ---
        & $\lesssim80$
        \\

        Si (E)\textsuperscript{f,*}
        \cite{Geyer2024NatPhys}
        & FinFET DQD
        & ---
        & ---
        & ---
        & $\sim31$
        \\

        Ge (E)\textsuperscript{g,*}
        \cite{Liu2022PRApplied}
        & Hut wire DQD
        & ---
        & ---
        & ---
        & $\sim2.0$
        \\

        Ge (E/T)\textsuperscript{h,\dag}
        \cite{Gao2020AdvMater}
        & Hut wire
        & ---
        & ---
        & ---
        & $\sim200$
        \\

        \bottomrule
        \bottomrule
    \end{tabular}

    \vspace{0.4em}

    \begin{minipage}{0.98\textwidth}
        \footnotesize\raggedright

        \textsuperscript{a}
        The slope is normalized to the internal field, $\mathcal{E}_{\mathrm{NW}}=2\mathcal{E}_{\mathrm{ext}}/17.8$, $\mathcal{E}_{\mathrm{ext}}\simeq10$--$20~\mathrm{V/\mu m}$. The maximum coefficient is read from Fig.~3(b) at $\mathcal{E}_{\mathrm{ext}}=80~\mathrm{V/\mu m}$.

        \par\smallskip

        \textsuperscript{b}
        Transport-fit estimate with $m^*=0.014m_\mathrm{e}$ and $\ell_{\mathrm{SO}}\simeq20~\mathrm{nm}$. Possible overestimation due to neglected orbital effects and intersubband coupling is discussed in Ref.~\cite{wojcik2018tuning}. The diameter refers to the electrostatic model.

        \par\smallskip

        \textsuperscript{c}
        The average internal field in the InAs experiment was inferred using the bulk InAs Rashba parameter.

        \par\smallskip

        \textsuperscript{d}
        Order-of-magnitude estimate based on the single-dot spin-orbit gap $\Delta_{\mathrm{SO}}\simeq110~\mu\mathrm{eV}$ extracted from a singlet--triplet anticrossing in Pauli-spin-blockade leakage.

        \par\smallskip

        \textsuperscript{e}
        Reported only as ``a few tens of nanometers,'' comparable to the interdot distance $d=80~\mathrm{nm}$.

        \par\smallskip

        \textsuperscript{f}
         Interdot distance $d\simeq40~\mathrm{nm}$. Consistent with the single-qubit EDSR estimate $\ell_{\mathrm{SO}}=20$--$60~\mathrm{nm}$ of Ref.~\cite{Camenzind2022NatElectron}.

        \par\smallskip

        \textsuperscript{g}
        Gate-tunable $\ell_{\mathrm{SO}}=2.0$--$48.9~\mathrm{nm}$ from Pauli-spin-blockade leakage, using $t_{\mathrm{SO}}/t=\tfrac{4}{3}l_{\mathrm{dot}}/\ell_{\mathrm{SO}}$ with $l_{\mathrm{dot}}\simeq75~\mathrm{nm}$. Since all fits give $t_{\mathrm{SO}}>t$, i.e., $\ell_{\mathrm{SO}}<l_{\mathrm{dot}}$, these values lie outside the perturbative regime of this relation; the shortest value is driven mainly by the suppression of $t$.

        \par\smallskip

        \textsuperscript{h}
        Range $200$--$600~\mathrm{nm}$ at $\mathcal{E}_z=-5~\mathrm{V/\mu m}$, from $\ell_{\mathrm{SO}}=\hbar^2/(m^*|\alpha'|)$ with the Rashba coefficient obtained from a singlet--triplet anticrossing and $m^*$ bounded by the Ge light- and heavy-hole masses.

        \par\smallskip

        \textsuperscript{*}
        For these entries, $\ell_{\mathrm{SO}}$ is inferred from the ratio $t_{\mathrm{SO}}/t$ of spin-flip to spin-conserving interdot tunnelling, obtained by fitting a two-site model to the measured cavity response (e), exchange anisotropy (f), or spin-blockade leakage current (g), and converted to $\ell_{\mathrm{SO}}$ using an assumed interdot distance. 
        In a two-site system this ratio is gauge dependent: spin-orbit-assisted tunnelling cannot be distinguished from a relative rotation of the $g$ tensors of the two dots. 
        The extracted $\ell_{\mathrm{SO}}$ therefore relies on a convention for the $g$ tensors.

        \par\smallskip

        \textsuperscript{\dag}
        Extracted from a single-dot two-hole singlet--triplet anticrossing. This avoids the interdot ambiguity of the DQD entries, but in a quasi-1D dot the anticrossing is equivalent to an intradot gradient of the $g$ tensor, so the extracted $\ell_{\mathrm{SO}}$ depends on the assumed effective mass and confinement model.

    \end{minipage}
\end{table*}

In the preceding analysis, we used the $\mathbf{k}\cdot\mathbf{p}$ parameters reported in Ref.~\cite{pulcu2024multiband}, which were obtained by fitting to the \emph{ab initio} band structure.

It is worth noting how accurately the individual parameters can be extracted from a fit to the band structure. 
Off-diagonal matrix elements influence the band energies in second order of perturbation theory, so an off-diagonal term of order $k^n$ contributes to the spectrum at order $k^{2n}$. 
The interband couplings $p_{\perp}$ and $p_\parallel$ enter the Hamiltonian linearly in $k$. Therefore, their contribution to the dispersion is of the same order in $k$ as that of the diagonal second-order terms: quadratic. As a result, the fit may partly trade off $p_\perp$ and $p_\parallel$ against the diagonal second-order parameters, leaving some freedom in their precise values. 
The parameters $A_5$ and $A_6$ multiply off-diagonal terms that are already quadratic in $k$, so they affect the spectrum at order $k^4$. At this order, their contribution overlaps with that of fourth-order diagonal terms, which lie beyond the second-order $\mathbf{k}\cdot\mathbf{p}$ model, and the fitted values may effectively include part of these higher-order effects. In contrast, $\Delta_1, \Delta_2$ and $\Delta_3$ are independent of $k$ and are fixed by the band energies at the $\Gamma$ point.

The question arises of how sensitive the Rashba coefficients obtained in the nanowire system are to the precise values of these parameters. To identify which interband couplings affect the Rashba splitting most, we set each of the four relevant $\mathbf{k}\cdot\mathbf{p}$ parameters, $p_\perp$, $A_5$, $A_6$, and the spin--orbit parameter $\Delta_3$, to zero one at a time. For each case we compute the relative change $\delta = |\Delta E - \Delta E^{(0)}|/\Delta E^{(0)}$ of the splitting defined in Eq.~\eqref{eq:linearenergysplitting}, where $\Delta E^{(0)}$ is the splitting with all parameters included. Since $\Delta E = 2\alpha_R|k_x|$ at fixed $k_x$, $\delta$ is also the relative change of the Rashba coefficient $\alpha_R$. The results for electric fields along $y$ and $z$ are summarized in Table~\ref{tab:rashba_params}.

For $\mathcal{E}_y$, the splitting is controlled mainly by $p_\perp$ and $A_5$: switching either one off reduces it by about $80\%$. $\Delta_3$ gives a moderate change of about $34\%$, and $A_6$ is almost irrelevant, at about $1\%$. For $\mathcal{E}_z$, the picture is different. Setting $\Delta_3 = 0$ removes the splitting completely ($\delta = 100\%$), so the Rashba coupling for $\mathcal{E}_z$ arises entirely through spin--orbit coupling. $p_\perp$ is also crucial ($\delta \approx 97\%$), while $A_5$ and $A_6$ contribute comparably and more weakly ($20\%$--$27\%$). The relative changes do not sum to $100\%$ in either field direction. This shows that the coupling channels do not contribute independently; splitting comes from their combined action. Overall, the Rashba splitting in the 2H-Ge nanowire depends strongly on both the field direction and the coupling channel, and only $p_\perp$ is essential for both directions.

The above analysis shows that the Rashba splitting is particularly sensitive to $p_\perp$ and, in the case of a $ y$-directional field, to $A_5$. The numerical values of $\alpha_R$ reported here should therefore be treated with caution because their accuracy is limited by that of the underlying $\mathbf{k}\cdot\mathbf{p}$ parameters. We also note that the off-diagonal parameters could be determined more precisely by directly using the DFT wavefunctions, which would lead to more accurate estimates of the Rashba parameters.

\section{Comparison with other semiconductor platforms}

\label{sec:rashba_comparison}
To place the electrically induced Rashba coupling in 2H-Ge in a broader context, Table~\ref{tab:rashba_comparison} compares selected semiconductor nanowire platforms, together with representative cubic Si and Ge hole quantum-dot devices. Alongside the field-normalized coefficients $\alpha_R'$, defined in Eq.~\eqref{eq:alpha_prime} and expressed in $e\,\mathrm{nm^2}$, the table lists the largest Rashba coefficients and shortest spin-orbit lengths reported in, or inferred from, the studies considered. These quantities characterize the specific devices and parameter ranges investigated rather than intrinsic material limits, as they depend on carrier species, confinement, subband structure, electrostatic conditions, and the extraction method.

The electric-field conventions used in these comparisons require particular care. The theoretical InSb coefficient is normalized to the screened field inside the nanowire~\cite{zhang2006rashba}, whereas the average internal field in the InAs experiment is inferred using the bulk InAs Rashba parameter~\cite{liang2012strong}. The theoretical Ge/Si coefficient describes the low-field direct Rashba spin-orbit interaction~\cite{kloeffel2018direct}. No field-normalized coefficient is assigned to the experimental InSb and Ge/Si entries or to the selected GaAs configuration; these entries instead provide comparisons of the reported coupling strengths and spin-orbit lengths. In particular, the large experimental InSb coefficient is inferred from a transport-model fit \cite{kammhuber2017conductance}, whose possible overestimation has been discussed in Ref.~\cite{wojcik2018tuning}.

For the $L=30~\mathrm{nm}$ 2H-Ge nanowire, we obtain 
\begin{align}
    \alpha_R^{\prime y}
    &\simeq 3.48~e\,\mathrm{nm^2}, &
    \alpha_R^{\prime z}
    &\simeq 0.81~e\,\mathrm{nm^2}.
\end{align}
For $\boldsymbol{\mathcal{E}}\perp c$, the field-normalized response is comparable to the theoretical InSb estimate ($\sim3.6~e\,\mathrm{nm^2}$), smaller than the Ge/Si direct-Rashba value ($\sim8.4~e\,\mathrm{nm^2}$), and larger than the inferred InAs value ($\sim1.17~e\,\mathrm{nm^2}$). For $\boldsymbol{\mathcal{E}}\parallel c$, the response is smaller than, but of the same order as, the InAs estimate. The two field orientations therefore differ by a factor of approximately $4.3$ in the weak-field regime, highlighting the pronounced directional anisotropy of electrical control in 2H-Ge nanowires. 

We also compare the spin-orbit length ($\ell_{\mathrm{SO}}$) values for different material systems in the sixth column of Table~\ref{tab:rashba_comparison}. For the Si and Ge quantum-dot devices in the lower block, only $\ell_{\mathrm{SO}}$ is available; it is extracted with device-specific definitions (see the table footnotes) and should be regarded as an order-of-magnitude reference.
The angular dependence of the electric-field response provides an additional means of enhancing the Rashba coupling.
At $\theta\simeq83.2^\circ$, the angular sweep yields a maximum Rashba coefficient of $\alpha_R^{\max}\simeq16.44~\mathrm{meV\,nm}$, with a corresponding minimum spin-orbit length of
$\ell_{\mathrm{SO}}^{\min}\simeq62.87~\mathrm{nm}$ for $\mathcal{E} = 7~\mathrm{V/\mu m}$. This value lies between those reported for Si FinFET ($\sim31~\mathrm{nm}$)~\cite{Geyer2024NatPhys} and planar Si ($\sim110~\mathrm{nm}$)~\cite{Li2015NanoLett} hole devices.

\section{Conclusion}

We have studied the magnetic- and electric-field response of confined hole states in hexagonal germanium nanowires using a ten-band $\mathbf{k}\cdot\mathbf{p}$ model. We find that the top bulk valence band has a vanishing $g$-factor in the plane transverse to the $ c$-axis, in close analogy to heavy holes in two-dimensional hole gases. In nanowires, confinement induces only small transverse components that vanish as the cross section grows; hence, for a generic magnetic-field direction, the qubit Zeeman splitting is controlled essentially by the field component along $c$.

We find that transverse electric field strongly reduces the nanowire band gap, providing a gate-tunable optical redshift and an upper bound on usable gate fields. The same field induces a Rashba coupling of the top valence subband that is considerably stronger for fields perpendicular to the $c$ axis and depends nonlinearly on field strength. In narrow wires at strong fields, the spin-orbit length reaches about $60$~nm, comparable to hole devices in cubic silicon, and the linear-in-momentum Rashba model remains accurate for gate-defined quantum dots of length $\approx 50$ nm throughout the electric-field range considered. In the weak-field regime, we identify that the Rashba coupling arises from second-order virtual processes through a few intermediate subbands, whose symmetry selection rules explain the directional anisotropy.

Together, the strongly anisotropic $g$ tensor, the gate-tunable Rashba interaction, and direct-gap optical activity make hexagonal germanium nanowires a promising group-IV platform for hole-spin qubits and for electrically and optically interfaced quantum devices. Our results, and the numerical workflow we developed here, account for key  parameters such as nanowire width,
crystal orientation relative to the gates, and the strength and direction of the gate field, and hence will prove useful to interpret and design future experiments.

\begin{acknowledgments}

We thank Z.~György, Y.~Pulcu, A.~Korm\'anyos, J.~Koltai, and G.~Burkard for valuable insights and helpful discussions, and our further partners in the ONCHIPS project for valuable feedback on the results.
This research is supported by the European Union within the Horizon Europe Research and Innovation Programme via the ONCHIPS project under Grant Agreement No. 101080022, and by the HUN-REN Hungarian Research Network through the Supported Research Groups Programme, HUN-REN-BME-BCE Quantum Technology Research Group (TKCS-2024/34). A.D. and D.V. acknowledge funding from the National Research, Development and Innovation Office of Hungary under OTKA grant no. FK 146499. D.V. was supported by the János Bolyai Research Scholarship of the Hungarian Academy of Sciences. B.K.~acknowledges support from the University Research Scholarship Program. Project no. EKÖP-26-4-I-BME-219 has been implemented with the support provided by the Ministry of Science and Technology of Hungary from the National Research, Development and Innovation Fund, financed under the EKÖP-26-4-I funding scheme.

\section*{Data availability}
The data and codes that support the findings of this article are openly available on Zenodo
at \textcolor{magenta}{\url{https://doi.org/10.5281/zenodo.23774497}}~\cite{Das2026_2HGe_code}.

\end{acknowledgments}

\section*{Author Contributions}
A.D. developed the numerical codes, performed the simulations, created the figures, and wrote the first draft of the manuscript. D.V., B.K., and A.D. collectively established the theoretical framework and carried out the symmetry analysis. All authors contributed significantly to the manuscript's revision and editing, as well as to critical discussions of the results. A.P. acquired funding, initiated the project, and supervised the work.


\appendix

\section{Matrix form of the k $\cdot$ p Hamiltonian}
\label{app:hamiltonian}

In this appendix section, we describe the \textbf{k}.\textbf{p} Hamiltonian that we have used for our numerical calculations.
The Hamiltonian representing the 10-band model of 2H-Ge can be written in matrix form \cite{Chunag96, pulcu2024multiband}:
\begin{widetext}
\begin{equation}
\resizebox{1.0\textwidth}{!}{$
\begin{aligned}
H_\text{\textbf{k}.\textbf{p}} &= 
\begin{pmatrix}
 C_2 + E_{cb'} & 0 & -\dfrac{\sqrt{2}\hbar k_+ p_{\perp}}{2m_\mathrm{e}} & \dfrac{\sqrt{2}\hbar k_- p_{\perp}}{2m_\mathrm{e}} & \dfrac{\hbar k_z p_{\parallel}}{m_\mathrm{e}} & 0 & 0 & 0 & 0 & 0 \\
0 &  C_1 + E_{cb} & 0 & 0 & 0 & 0 & 0 & 0 & 0 & 0 \\
-\dfrac{\sqrt{2}\hbar k_- p_{\perp}}{2m_\mathrm{e}} & 0 &  \Delta_1 + \Delta_2 + E_v + \alpha + \beta & +A_5 k_-^2 & +A_6 k_- k_z & 0 & 0 & 0 & 0 & 0 \\[1.2em]
\dfrac{\sqrt{2}\hbar k_+ p_{\perp}}{2m_\mathrm{e}} & 0 & +A_5 k_+^2 &  \Delta_1 - \Delta_2 + E_v + \alpha + \beta & -A_6 k_+ k_z & 0 & 0 & 0 & 0 & \sqrt{2}\Delta_3 \\
\dfrac{\hbar k_z p_{\parallel}}{m_\mathrm{e}} & 0 & +A_6 k_+ k_z & -A_6 k_- k_z &  E_v + \beta & 0 & 0 & 0 & \sqrt{2}\Delta_3 & 0 \\
0 & 0 & 0 & 0 & 0 &  C_2 + E_{cb'} & 0 & \dfrac{\sqrt{2}\hbar k_- p_{\perp}}{2m_\mathrm{e}} & -\dfrac{\sqrt{2}\hbar k_+ p_{\perp}}{2m_\mathrm{e}} & \dfrac{\hbar k_z p_{\parallel}}{m_\mathrm{e}} \\
0 & 0 & 0 & 0 & 0 & 0 &  C_1 + E_{cb} & 0 & 0 & 0 \\
0 & 0 & 0 & 0 & 0 & \dfrac{\sqrt{2}\hbar k_+ p_{\perp}}{2m_\mathrm{e}} & 0 &  \Delta_1 + \Delta_2 + E_v + \alpha + \beta & +A_5 k_+^2 & -A_6 k_+ k_z \\
0 & 0 & 0 & 0 & \sqrt{2}\Delta_3 & -\dfrac{\sqrt{2}\hbar k_- p_{\perp}}{2m_\mathrm{e}} & 0 & +A_5 k_-^2 &  \Delta_1 - \Delta_2 + E_v + \alpha + \beta & +A_6 k_- k_z \\
0 & 0 & 0 & \sqrt{2}\Delta_3 & 0 & \dfrac{\hbar k_z p_{\parallel}}{m_\mathrm{e}} & 0 & -A_6 k_- k_z & +A_6 k_+ k_z &  E_v + \beta
\end{pmatrix}
\end{aligned}
$}
\label{eq:hamiltonian}
\end{equation}
\end{widetext}
The ten-band Hamiltonian in Eq. \eqref{eq:hamiltonian} is expressed in the basis listed in Table~\ref{tab:basis}, which differs from the basis used in~\cite{pulcu2024multiband} in the relative phases of certain basis functions.

\begin{table}[t]
\caption{Basis functions for the $D_{6h}$ point group adapted from~\cite{Chunag96,pulcu2024multiband,Claudia2019accurate}.
The molecular orbitals $|S_x\rangle$ and $|S_y\rangle$ transform as axial vectors, $|Z\rangle$ transforms as a polar vector and $|\mathbbm{1}\rangle$
transforms trivially.
Spins transform according to the standard spin-$1/2$ representation.
We also list the corresponding single and double group representations of the $D_{6h}$ group.}
\label{tab:basis}
\begin{ruledtabular}
\begin{tabular}{crr}
Basis functions &  irrep. & double irrep. \\
\hline
$|Z\!\uparrow\rangle$                                        & $\Gamma_2^-$ & $\Gamma_7^-$ \\
$|\Gamma_3^-\!\uparrow\rangle$                               & $\Gamma_3^-$ & $\Gamma_8^-$ \\
$\frac{1}{\sqrt{2}}|S_x + iS_y\!\uparrow\rangle$            & $\Gamma_5^+$ & $\Gamma_9^+$ \\
$\frac{1}{\sqrt{2}}|S_x - iS_y\!\uparrow\rangle$             & $\Gamma_5^+$ & $\Gamma_7^+$ \\
$-i |\mathbbm{1}\!\uparrow\rangle$                              & $\Gamma_1^+$ & $\Gamma_7^+$ \\
$|Z\!\downarrow\rangle$                                     & $\Gamma_2^-$ & $\Gamma_7^-$ \\
$|\Gamma_3^-\!\downarrow\rangle$                             & $\Gamma_3^-$ & $\Gamma_8^-$ \\
$\frac{1}{\sqrt{2}}|S_x - iS_y\!\downarrow\rangle$           & $\Gamma_5^+$ & $\Gamma_9^+$ \\
$\frac{1}{\sqrt{2}}|S_x + iS_y\!\downarrow\rangle$          & $\Gamma_5^+$ & $\Gamma_7^+$ \\
$- i |\mathbbm{1}\!\downarrow\rangle$                             & $\Gamma_1^+$ & $\Gamma_7^+$ \\
\end{tabular}
\end{ruledtabular}
\end{table}

In Eq.~\eqref{eq:hamiltonian}, $\hbar$ is the reduced Planck constant, and $m_\mathrm{e}$ is the free-electron mass.
The optical transition matrix elements $p_{\parallel}$ and $p_{\perp}$ are defined as \cite{Chunag96, pulcu2024multiband}
\begin{subequations}
\begin{align}
    p_\parallel &= -i\bra{Z}p_z\ket{\mathbbm{1}},\\
    p_\perp &= -i\bra{Z}p_x\ket{S_y} = i\bra{Z}p_y\ket{S_x}.
\end{align} 
\end{subequations}
The in-plane momentum combinations are defined as $k_{\pm}=k_x\pm i k_y$. The remaining momentum-dependent quantities are defined as
\begin{align}
C_1 &=
A_{c1\perp}\left(k_x^2+k_y^2\right)
+A_{c1\parallel}k_z^2,
\\
C_2 &=
A_{c2\perp}\left(k_x^2+k_y^2\right)
+A_{c2\parallel}k_z^2,
\\
\alpha &=
A_3k_z^2+A_4\left(k_x^2+k_y^2\right),
\\
\beta &=
A_1k_z^2+A_2\left(k_x^2+k_y^2\right).
\end{align}
The numerical values of the optical-transition matrix elements, effective mass parameters, and energy splittings are summarized in Table~\ref{tab:kp_parameters}. 

Note, the Hamiltonian $H_{\mathbf{k}\cdot\mathbf{p}}$ used here differs from the explicit matrix representation reported in Ref.~\cite{pulcu2024multiband}. There are two points concerning its implementation, which we discuss below.

First, we have noticed a typo in the caption of Table~II of Ref.~\cite{pulcu2024multiband}, particularly in the definition of the $P_1$ and $P_2$ parameters. The $P_1$ parameter correctly reads as
\begin{equation}
    P_1 = -i\frac{\hbar}{m_\mathrm{e}}\bra{\Gamma_{2c}^-}p_z\ket{\Gamma_{1v}^+} = \frac{\hbar}{m_\mathrm{e}}p_\parallel,
\end{equation}
without a $-i$ factor for the last equation; otherwise, the Hamiltonian would not be hermitian. 
Similarly, the $P_2$ parameter in Ref.~\cite{pulcu2024multiband} correctly reads as
\begin{equation}
    P_2 = -i\frac{\hbar}{m_\mathrm{e}}\bra{\Gamma_{2c}^-}p_x\ket{S_y} = \frac{\hbar}{m_\mathrm{e}}p_\perp,
\end{equation}
without the $-i$ factor in the last equation. 

Second, and more importantly, the Hamiltonian does not appear to fully respect the symmetry of the hexagonal crystal structure after the addition of the linear term in Eq.~(4) and the second-order term in Eq.~(B5) of Ref.~\cite{pulcu2024multiband}. In particular, the mirror symmetries and $C_2$ rotations with axes in the $xy$ plane are broken~\cite{Varjas_2018}.

To restore the symmetry and find the proper form of the second-order terms in our basis, we take a step back. 
The second-order contributions, besides the free electron term, arise from the perturbative coupling to the distant bands (L\"owdin partitioning).
The conduction-band part of the second-order Hamiltonian is diagonal and therefore respects the symmetry; its form is unchanged with respect to Ref.~\cite{pulcu2024multiband}, since diagonal entries do not depend on the phases of the basis states. We therefore focus on the valence-band sector. 
In the real orbital basis $|S_x\rangle,|S_y\rangle,|\mathbbm{1}\rangle$, the second-order terms take the form
\begin{widetext}
\begin{equation}\label{eq:Dform}
D_{3\times3}=
\frac{\hbar^2}{2m_\mathrm{e}}\begin{pmatrix}
L_1k_x^2+M_1k_y^2+M_2k_z^2 & N_1k_xk_y & -N_2k_yk_z\\
N_1k_xk_y & M_1k_x^2+L_1k_y^2+M_2k_z^2 & N_2k_xk_z\\
-N_2k_yk_z & N_2k_xk_z & M_3(k_x^2+k_y^2)+L_2k_z^2
\end{pmatrix},
\end{equation}
with parameters
\begin{subequations}
\label{eqs:second_order_params}
\begin{align}
L_1&=1+\frac{2}{m_\mathrm{e}}\sum_\gamma\frac{\langle S_x|p_x|\gamma\rangle\langle\gamma|p_x|S_x\rangle}{E_v+\Delta_1-E_\gamma},
&M_1&=1+\frac{2}{m_\mathrm{e}}\sum_\gamma\frac{\langle S_x|p_y|\gamma\rangle\langle\gamma|p_y|S_x\rangle}{E_v+\Delta_1-E_\gamma}, \\
M_2&=1+\frac{2}{m_\mathrm{e}}\sum_\gamma\frac{\langle S_x|p_z|\gamma\rangle\langle\gamma|p_z|S_x\rangle}{E_v+\Delta_1-E_\gamma},
&M_3&=1+\frac{2}{m_\mathrm{e}}\sum_\gamma\frac{\langle \mathbbm{1}|p_x|\gamma\rangle\langle\gamma|p_x|\mathbbm{1}\rangle}{E_v-E_\gamma}, \\
L_2&=1+\frac{2}{m_\mathrm{e}}\sum_\gamma\frac{\langle \mathbbm{1}|p_z|\gamma\rangle\langle\gamma|p_z|\mathbbm{1}\rangle}{E_v-E_\gamma},
\end{align}
\begin{align}
N_1&=\frac{2}{m_\mathrm{e}}\sum_\gamma\frac{\langle S_x|p_x|\gamma\rangle\langle\gamma|p_y|S_y\rangle+\langle S_x|p_y|\gamma\rangle\langle\gamma|p_x|S_y\rangle}{E_v+\Delta_1-E_\gamma}, \\
N_2&=\frac{1}{m_\mathrm{e}}\sum_\gamma\Big[\langle S_y|p_x|\gamma\rangle\langle\gamma|p_z|\mathbbm{1}\rangle+\langle S_y|p_z|\gamma\rangle\langle\gamma|p_x|\mathbbm{1}\rangle\Big] \times \left(\frac{1}{E_v+\Delta_1-E_\gamma}+\frac{1}{E_v-E_\gamma}\right).
\end{align}
\end{subequations}
\end{widetext}
In Eqs.~\eqref{eqs:second_order_params}, $\gamma$ runs over the distant bands.
The form of $D_{3\times 3}$ in Eq.~\eqref{eq:Dform} differs from Eq.~(B3) of Ref.~\cite{pulcu2024multiband} by the entries $(1,3)$, $(2,3)$ and $(3,1)$, $(3,2)$, which couple $|S_x\rangle$ and $|S_y\rangle$ to $|\mathbbm{1}\rangle$.
With the form of the second order terms in Eq.~(B3) in Ref.~\cite{pulcu2024multiband}, the Hamiltonian does not respect the $y \to -y$ mirror symmetry, because the basis functions $|S_x\rangle$ and $|S_y\rangle$ transform as axial vectors not as polar vectors.
With the form above it satisfies all point-group operations.

In the basis $|S_\pm\rangle=(|S_x\rangle\pm i|S_y\rangle)/\sqrt2$, $-i|\mathbbm{1}\rangle$ the valence-band block becomes
{\setlength{\arraycolsep}{1pt}\small
\begin{equation}
D=
\begin{pmatrix}
\frac{L_1+M_1}{2}k_\perp^2+M_2k_z^2 & \frac{N_1}{2}k_-^2 & \frac{N_2}{\sqrt2}k_-k_z\\[2pt]
\frac{N_1}{2}k_+^2 & \frac{L_1+M_1}{2}k_\perp^2+M_2k_z^2 & -\frac{N_2}{\sqrt2}k_+k_z\\[2pt]
\frac{N_2}{\sqrt2}k_+k_z & -\frac{N_2}{\sqrt2}k_-k_z & M_3k_\perp^2+L_2k_z^2
\end{pmatrix}.
\end{equation}
}
For the spin-down block the order of the orbital wavefunctions $\ket{S_+}$ and $\ket{S_+}$ is changed, therefore the matrix elements should be permuted in the spin-down block accordingly.

Introducing the $A$ parameters \cite{Chunag96}:
\begin{equation}
\begin{alignedat}{2}
A_1&=L_2, 
& A_2&=M_3, \\
A_3&=M_2-L_2, \qquad
&A_4&=\tfrac{L_1+M_1}{2}-M_3, \\
A_5&=\tfrac{N_1}{2}, 
&A_6&=\tfrac{N_2}{\sqrt2},
\end{alignedat}
\end{equation}
we obtain exactly the second-order entries of the Hamiltonian in Eq.~\eqref{eq:hamiltonian} after substituting $\alpha=A_3k_z^2+A_4k_\perp^2$ and $\beta=A_1k_z^2+A_2k_\perp^2$.
The Hamiltonian is therefore written in a single consistent basis (Table~\ref{tab:basis}), in which the linear and quadratic terms follow from the same representation and respect time reversal and the point group symmetries.

\begin{table}[t]
\caption{
Parameters of the ten-band $\mathbf{k}\cdot\mathbf{p}$ Hamiltonian for 2H-Ge. The optical-transition matrix elements are given in units of $\hbar/a_0$, the effective mass parameters in units of $\hbar^2/(2m_\mathrm{e})$, and the energy splittings in eV. The parameter values are taken from Ref.~\cite{pulcu2024multiband}; further details of their derivation and fitting procedure can be found therein. }
\label{tab:kp_parameters}
\centering
\begin{ruledtabular}
\begin{tabular}{lc}
Parameter & $H_{10\times 10}^{\mathrm{hex\text{-}Ge}}$ \\
\colrule

\multicolumn{2}{l}{Optical-transition matrix elements} \\
$p_{\perp}$     & $0.4829$ \\
$p_{\parallel}$ & $0.6431$ \\

\multicolumn{2}{l}{Conduction-band effective mass parameters} \\
$A_{c2\perp}$     & $4.1565$ \\
$A_{c1\perp}$     & $9.5120$ \\
$A_{c2\parallel}$ & $2.4091$ \\
$A_{c1\parallel}$ & $1.0$ \\

\multicolumn{2}{l}{Valence-band effective mass parameters} \\
$A_1$ & $-4.3636$ \\
$A_2$ & $-2.0833$ \\
$A_3$ & $2.4545$ \\
$A_4$ & $-2.7504$ \\
$A_5$ & $ 2.7232$ \\
$A_6$ & $-3.5421$ \\

\multicolumn{2}{l}{Energy splittings} \\
$\Delta_1=\Delta_{\mathrm{cf}}$              & $0.2688$ \\
$\Delta_2=\Delta_{\mathrm{so}}^{\parallel}/3$ & $0.0934$ \\
$\Delta_3=\Delta_{\mathrm{so}}^{\perp}/3$     & $0.0908$ \\

\multicolumn{2}{l}{Other energy parameters} \\
$E_{cb}$              & $0.298$ \\
$E_{cb'}$              & $0.632$ \\
$E_{v}$              & $-0.3622$ \\

\end{tabular}
\end{ruledtabular}
\end{table}

\section{Effective mass calculation}\label{app:effectivemass}
In this appendix, we first calculate the effective masses of the top bulk valence band from the $\mathbf{k}\cdot\mathbf{p}$ Hamiltonian expanded around the $\Gamma$ point (Sec.~\ref{app:effmass}). We then compute how the longitudinal effective mass changes due to the finite cross section of the nanowire (Sec.~\ref{app:effmass_nw}). \subsection{Effective mass of the top bulk valence band}
\label{app:effmass}

The effective mass of the top bulk valence band can be obtained by expanding the bulk $\mathbf{k}\cdot\mathbf{p}$ Hamiltonian around the $\Gamma$ point. We write the Hamiltonian as
\begin{align}
H_{\mathbf{k}\cdot\mathbf{p}}
&=
H^{(0)}
+
H^{(1)}(\mathbf{k})
+
H^{(2)}(\mathbf{k}),
\end{align}
where $H^{(0)}$ is the Hamiltonian at $\mathbf{k}=0$, $H^{(1)}$ contains terms linear in momentum, and $H^{(2)}$ contains terms quadratic in momentum. In the basis of Eq.~(A1), the top bulk-valence-band Kramers pair is
\begin{subequations}
\begin{align}
|\mathrm{VB}_1,\uparrow\rangle &\equiv \frac{1}{\sqrt{2}}|S_x +iS_y, \uparrow\rangle,
\\
|\mathrm{VB}_1,\downarrow\rangle &\equiv \frac{1}{\sqrt{2}}|S_x -iS_y, \downarrow\rangle,
\end{align}    
\end{subequations}
with energy
\begin{align}
E_{\mathrm{VB}_1} = E_v+\Delta_1+\Delta_2 .
\end{align}

We define the projection operator onto the complementary subspace as
\begin{align}
Q = \mathbb{I} -(
|\mathrm{VB}_1,\uparrow\rangle\langle \mathrm{VB}_1,\uparrow|
+
|\mathrm{VB}_1,\downarrow\rangle\langle \mathrm{VB}_1,\downarrow|),
\end{align}
and the isometry
\begin{equation}
    V_{\mathrm{VB}_1} = |\Uparrow\rangle\langle \mathrm{VB}_1,\uparrow|
+
|\Downarrow\rangle\langle \mathrm{VB}_1,\downarrow|,
\end{equation}
where $|\Uparrow\rangle$ and $|\Downarrow\rangle$ span the two-dimensional Hilbert space corresponding to the single-band effective mass model of the top valence band.
The isometry $V_{\mathrm{VB}_1}$ projects the Hamiltonian to the two-dimensional subspace corresponding to the top valence band and then truncates the Hilbert space.

Using Schrieffer--Wolff perturbation theory, the effective Hamiltonian in the top-valence-band subspace, up to second order in momentum, is 
\begin{align}\label{eq:VB1SW}
\begin{split}
H_{\mathrm{eff}}^{\mathrm{VB}_1}
&=
E_{\mathrm{VB}_1}\mathbb{I}_{2\times 2}
+
V_{\mathrm{VB}_1} H^{(2)} V_{\mathrm{VB}_1}^\dagger\\
&\quad+
V_{\mathrm{VB}_1}H^{(1)} Q
\frac{1}{E_{\mathrm{VB}_1}-QH^{(0)}Q}
Q H^{(1)}V_{\mathrm{VB}_1}^\dagger.
\end{split}
\end{align}

Since $\langle\mathrm{VB}_1,\uparrow|H^{(2)}|\mathrm{VB}_1\downarrow\rangle = 0$, the direct quadratic contribution is diagonal; moreover it is proportional to the two-by-two identity and reads as
\begin{equation}
    V_{\mathrm{VB}_1} H^{(2)} V_{\mathrm{VB}_1}^\dagger = (A_2+A_4)(k_x^2+k_y^2) \mathbb{I}_{2\times 2}
    +
    (A_1+A_3)k_z^2 \mathbb{I}_{2\times 2} .
\end{equation}

The top valence band only couples to the second conduction band through the linear-in-momentum Hamiltonian term $H^{(1)}$, and it does not mix the spin-up and spin-down sectors; therefore, the contribution of the third term in Eq.~\eqref{eq:VB1SW} reads as
\begin{multline}
    V_{\mathrm{VB}_1} H^{(1)} Q
    \frac{1}{E_{\mathrm{VB}_1}-QH^{(0)}Q} Q H^{(1)} V_{\mathrm{VB}_1}^\dagger = 
    \\
    =\frac{\hbar^2p_\perp^2}{2m_\mathrm{e}^2(E_\mathrm{VB_1} - E_\mathrm{CB_2})}(k_x^2 + k_y^2) \mathbb{I}_{2\times 2}
\end{multline}

Hence, up to second order in momentum, the effective Hamiltonian of the top bulk-valence-band doublet becomes
\begin{align}
H_{\mathrm{eff}}^{\mathrm{VB}_1}
&=
\left[
E_{\mathrm{VB}_1}
+
a_\perp(k_x^2+k_y^2)
+
a_z k_z^2
\right]\mathbb{I}_{2\times 2},
\end{align}
where
\begin{align}
a_\perp
&=
(A_2+A_4)
+
\frac{\hbar^2p_\perp^2}{2m_\mathrm{e}^2(E_{\mathrm{VB}_1}-E_{\mathrm{CB}_2})},\label{eq:aperp}
\\
a_z
&=
A_1+A_3 . \label{eq:az}
\end{align}
The corresponding dispersion of the top bulk valence band near the $\Gamma$
point is therefore
\begin{align} \label{eq:VBdispersion}
E_{\mathrm{VB}_1}(\mathbf{k})
&=
E_{\mathrm{VB}_1}
+
a_\perp(k_x^2+k_y^2)
+
a_z k_z^2 .
\end{align}

\subsubsection{Effective mass perpendicular to the crystal $c$ axis, $m_x=m_y$}

From the quadratic dispersion in Eq.~\eqref{eq:VBdispersion}, the effective mass perpendicular to the crystal
$c$ axis is
\begin{align}
\frac{1}{m_x}
&=
\frac{1}{\hbar^2}
\frac{\partial^2 E_{\mathrm{VB}_1}}{\partial k_x^2}
=
\frac{2a_\perp}{\hbar^2}.
\end{align}
Hence from Eq.~\eqref{eq:aperp},
\begin{align}
m_x
&=
\frac{\hbar^2 m_\mathrm{e}^2 (E_{\mathrm{VB}_1}-E_{\mathrm{CB}_2})}
{
2m_\mathrm{e}^2 (E_{\mathrm{VB}_1}-E_{\mathrm{CB}_2})(A_2+A_4)
+
\hbar^2 p_\perp^2
}.
\end{align}
Using the parameter values from Table~\ref{tab:kp_parameters}, we obtain
\begin{align}
m_x
&=
-0.0672\,m_\mathrm{e} .
\end{align}
Here, $m_\mathrm{e}$ is the free-electron mass. The negative sign reflects the downward curvature of the valence band.

\subsubsection{Effective mass parallel to the crystal $c$ axis, $m_z$}

From the quadratic dispersion in Eq.~\eqref{eq:VBdispersion}, the effective mass parallel to the crystal $c$-axis is
\begin{align}
\frac{1}{m_z}
&=
\frac{1}{\hbar^2}
\frac{\partial^2 E_{\mathrm{VB}_1}}{\partial k_z^2}
=
\frac{2a_z}{\hbar^2}.
\end{align}
Hence from Eq.~\eqref{eq:az},
\begin{align}
m_z
&=
\frac{\hbar^2}{2a_z}
=
\frac{\hbar^2}{2(A_1+A_3)} .
\end{align}
Using the parameter values from Table~\ref{tab:kp_parameters}, we obtain
\begin{align}
m_z
&=
-0.524\,m_\mathrm{e} .
\end{align}
Again, the negative sign reflects the downward curvature of the valence band.

\subsection{Confinement dependence of the longitudinal effective mass}
\label{app:effmass_nw}

In addition to the bulk effective mass estimate, we also extract the longitudinal effective mass directly from the confined nanowire Hamiltonian. For each cross-sectional side length $L$, we diagonalize the mixed-space nanowire Hamiltonian $H_0(k_x)$ and identify the top confined valence subband $\mathrm{vb}_0$. We then use a Schrieffer--Wolff projection, implemented with \textsc{Pymablock}, to obtain the effective low-energy Hamiltonian of this subband near $k_x=0$. The resulting dispersion is fitted to the form
\begin{align}
E_{\mathrm{vb}_0}(k_x)
=
E_{\mathrm{vb}_0}(0)
+
\frac{\hbar^2 k_x^2}{2m_x^*(L)}
\end{align}
from which the longitudinal nanowire effective mass $m_x^*(L)$ is extracted.

\begin{figure}[t]
    \centering
    \includegraphics[width=.8\columnwidth]{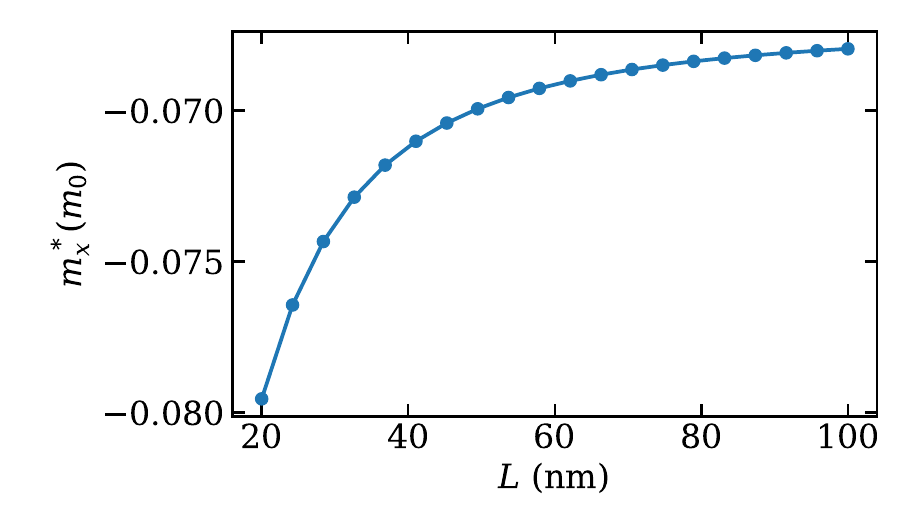}
    \caption{
    Cross-sectional size ($L$) dependence of the longitudinal effective mass $m_x^*$ of the top valence subband.
    }
    \label{fig:effective_mass_modulation}
\end{figure}

Fig.~\ref{fig:effective_mass_modulation} shows the resulting dependence of $m_x^*$ on the nanowire cross-sectional side length. The effective mass remains negative, as expected for a valence-band-derived subband, but its magnitude decreases as $L$ increases. This trend indicates that confinement-induced band mixing is strongest for smaller cross sections and gradually weakens as the nanowire approaches the bulk limit.

\section{Modification of hoppings due to an external magnetic field}
\label{app:peierls}
In this appendix, we describe how the tight-binding model introduced in Eq.~\eqref{eq:H0} is modified in the presence of an external magnetic field, which enters through Peierls phases.

We label each transverse site $i$ by its integer grid coordinates $(m_i,n_i)$, corresponding to the position $(y_i,z_i)=(m_i\Delta,n_i\Delta)$, where $\Delta$ is the
discretization length and the origin lies on the nanowire axis.

A magnetic field is included through the Peierls substitution: every real-space hopping between lattice sites at $\mathbf{r}_i$
and $\mathbf{r}_j$ acquires the factor 
\begin{equation}
    \exp\!\left(i\frac{2\pi}{\phi_0}\int_{\mathbf{r}_i}^{\mathbf{r}_j}
    \mathbf{A}\cdot d\mathbf{l}\right),
    \label{eq:peierls}
\end{equation}
where $\phi_0 = h/e$ is the flux quantum and the integral is taken along the straight bond. This convention corresponds to the minimal coupling $\mathbf{k}\rightarrow\mathbf{k}+\tfrac{2\pi}{\phi_0}\mathbf{A}$.
In all gauges used below, $\mathbf{A}$ is independent of $x$ and linear in the transverse coordinates. Translational invariance along the wire is therefore preserved, $k_x$ remains a good quantum number, and the integral in Eq.~\eqref{eq:peierls} is given exactly by the vector potential at the bond center. We denote the magnetic flux through one plaquette of the transverse grid by
$\Phi = B\Delta^2/\phi_0$.

\subsection{Longitudinal magnetic field}
For a field along the wire, $\mathbf{B}=B\hat{x}$, we choose the gauge $\mathbf{A}=(0,0,By)$. The Peierls phase of a bond then
depends only on its $z$-displacement and on the $y$-coordinate of its center, and not on its displacement along $x$. It therefore
multiplies the full $k_x$-dependent hopping matrix:
\begin{equation}
    t_{ab}^{ij,k_x} \rightarrow t_{ab}^{ij,k_x}\,
    e^{i\pi\Phi\,(m_i+m_j)(n_j-n_i)}.
    \label{eq:peierls_long}
\end{equation}
On-site terms and hoppings along $y$ ($n_i=n_j$) are unaffected,
whereas hoppings along $z$ and diagonal hoppings acquire a phase
proportional to the $y$-coordinate of the bond center.

\subsection{Transverse magnetic field}

For a field along $z$, $\mathbf{B}=B\hat{z}$, we choose the gauge
$\mathbf{A}=(-By,0,0)$. Since $\mathbf{A}$ is parallel to the wire,
bonds lying in the transverse plane acquire no phase. A bond with
displacement $\Delta x$ along the wire acquires the phase
$-\tfrac{2\pi}{\phi_0}B\,\tfrac{y_i+y_j}{2}\,\Delta x$, which
combines with the Bloch factor $e^{ik_x\Delta x}$. The field thus
replaces $k_x$ by its minimally coupled value with the vector
potential evaluated at the bond center:
\begin{equation}
    t_{ab}^{ij,k_x} \rightarrow t_{ab}^{ij,k_x^{ij}},
    \qquad
    k_x^{ij} = k_x - \frac{\pi\Phi}{\Delta}\,(m_i+m_j).
    \label{eq:peierls_trans}
\end{equation}
This includes the on-site terms, $i=j$, for which
$k_x^{ii} = k_x - 2\pi\Phi m_i/\Delta$. In contrast to the
longitudinal case, the field does not multiply the hopping matrices
by a phase but shifts their $k_x$ argument. 

For a field along $y$, $\mathbf{B}=B\hat{y}$, the analogous gauge
$\mathbf{A}=(Bz,0,0)$ leads to Eq.~\eqref{eq:peierls_trans} with
\begin{equation}
    k_x^{ij} = k_x + \frac{\pi\Phi}{\Delta}\,(n_i+n_j).
\end{equation}

\section{Vanishing of the bulk transverse $g$-factors}
\label{app:bulk_xy_gfactor}
In this appendix we prove that the $g_{xx}$ and $g_{yy}$ $g$-factors vanish for the top valence band of bulk hexagonal germanium.
Consider the generic bulk Zeeman term restricted to the top valence band up to first order in the magnetic field:
\begin{equation}
    H_Z (\mathbf{B}) = \sum_{ij} c_{ij} \sigma_i B_j.
\end{equation}
We write the Hamiltonian $H_Z$ in the basis of the $\Gamma$ point $\mathrm{VB}_1$ basis functions $\frac{1}{\sqrt{2}}|S_x + iS_y\!\uparrow\rangle$ and $\frac{1}{\sqrt{2}}|S_x - iS_y\!\downarrow\rangle$, $\sigma_i$ denote the Pauli matrices in this basis.
An effective Hamiltonian of this form can be obtained by perturbation theory at $\mathbf{k} = 0$ including both the spin Zeeman term and the orbital magnetic field.
The only property we will use is the symmetry transformation properties of the $\mathrm{VB}_1$ states (double irrep $\Gamma_9^+$, see Table~\ref{tab:basis}), which is maintained at all orders of perturbation theory.

The sixfold rotational symmetry of the system imposes the constraint~\cite{Miles_2025}
\begin{equation}
    U_{C_{6z}} H_Z (C_{6z}^{-1}\mathbf{B}) U_{C_{6z}}^{\dagger} = H_Z(\mathbf{B}),
\end{equation}
where $C_{6z}$ is the real-space sixfold rotation matrix, and $U_{C_{6z}} = -i \sigma_z$ is its representation in the $\Gamma_9^+$ irrep.
Substituting the generic form above gives, after straightforward calculation, that $c_{ij} = 0$ except for $i = j = z$.
The only symmetry-allowed term in $H_Z$ is proportional to $B_z \sigma_z$, meaning only a $B_z$ field can cause a splitting, corresponding to $ z$-spin polarization; hence, $g_{zz}$ is the only nonzero component of the $g$-tensor.

Another way of arriving at this result is to consider the representation content of the Hamiltonian terms under the $D_{6h}$ symmetry group of the crystal.
The $2 \times 2$ matrix part of the Hamiltonian transforms according to the representation $\Gamma_9^+ \otimes (\Gamma_9^+)^* = \Gamma_1^+ \oplus \Gamma_2^+ \oplus \Gamma_3^+ \oplus \Gamma_4^+$, where we used the character table to decompose the representation.
The magnetic field vector $\mathbf{B}$ transforms as $\Gamma_2^+ \oplus \Gamma_5^+$, where the first component is a 1D irrep corresponding to the $B_z$ component and the second is a 2D irrep corresponding to the $(B_x, B_y)$ components transforming into each other.
The generic Hamiltonian $H_Z$ is composed of a matrix times $\mathbf{B}$ bilinear, and carries the product representation of the two above.
However, the symmetric part needs to be invariant, i.e. transform under the trivial representation.
The trivial representation can only appear as the product of an irrep with its own complex conjugate, and all $D_{6h}$ irreps are equivalent to their complex conjugates.
From this we see that the invariant Hamiltonians form the 1D subspace $\Gamma_2^+ \otimes \Gamma_2^+$ that only includes the $\Gamma_2^+$ part of both representations, corresponding exactly to constant multiples of $\sigma_z B_z$.

\section{The $g_{zz}$ element of the $g$ tensor from perturbation theory in the envelope-function model}
\label{app:bulkgfactor}

In this appendix, we derive $g_{zz}$, the only nonzero element of the bulk $g$ tensor of the top valence band $\mathrm{VB}_1$, see Appendix \ref{app:bulk_xy_gfactor}.
We consider a nanowire oriented along $x$ with a square cross-section of side length $L$ in the $(y,z)$ plane.
We construct the Kramers pair of the highest confined valence subband $\mathrm{vb}_0$, which originates predominantly from the bulk band $\mathrm{VB}_1$, perturbatively for large $L$, compute its Zeeman splitting for a magnetic field along the $z$ axis, and recover the bulk result in the limit $L\to\infty$.

\subsection{Unperturbed problem}

The unperturbed Hamiltonian consists of the diagonal part of the $\mathbf{k}\cdot\mathbf{p}$ Hamiltonian in Eq.~\eqref{eq:hamiltonian} and the transverse hard-wall potential
\begin{equation}
    V(y, z) =
    \begin{cases}
       0,      & (y,z) \in [-L/2, L/2]^2, \\
       \infty, & \text{otherwise}.
    \end{cases}
\end{equation}
The off-diagonal $\mathbf{k}\cdot\mathbf{p}$ terms are treated as a perturbation.
In the envelope-function approximation, the unperturbed eigenstates are
\begin{equation}
    \ket{b; n_y, n_z} = \ket{u_b}\,\phi_{n_y}(y)\, \phi_{n_z}(z),
\end{equation}
where $\ket{u_b}$ is the bulk Bloch state of band $b$ at the $\Gamma$ point (see Table~\ref{tab:basis}) and $\phi_{n}$ are the eigenstates of the one-dimensional hard-wall well,
\begin{equation}
    \phi_n(w)
    =
    \sqrt{\frac{2}{L}}
    \begin{cases}
        \cos\left(\dfrac{n\pi w}{L}\right),
        & n \text{ odd},\\[6pt]
        \sin\left(\dfrac{n\pi w}{L}\right),
        & n \text{ even},
    \end{cases}
    \qquad n = 1, 2, \dots
    \label{eq:box_eigenstates}
\end{equation}
The corresponding zeroth-order energies are
\begin{equation}\label{eq:2Dbox_spectrum}
    E_{b;n_y,n_z}^{(0)} = E_b + \frac{\pi^2\hbar^2}{2L^2}
    \left(
    \frac{n_y^2}{m_{b,y}^*}
    +
    \frac{n_z^2}{m_{b,z}^*}
    \right),
\end{equation}
where $E_b$ is the energy of band $b$ at the $\Gamma$ point and the mass parameters $m^*_{b,i}$ are obtained from the diagonal elements of Eq.~\eqref{eq:hamiltonian} as
\begin{equation}
    \frac{1}{m^*_{b,i}} = \frac{1}{\hbar^2}\,
    \frac{\partial^2}{\partial k_i^2}\bra{u_b}H_{\mathbf{k}\cdot\mathbf{p}}(\mathbf{k})\ket{u_b}\Big|_{\mathbf{k}=0}.
\end{equation}
Note that $m^*_{b, i}$ is not the bulk effective mass of band $b$, because the off-diagonal terms are neglected.
For the valence band $\mathrm{VB}_1$, these mass parameters are negative, so that the state with $n_y=n_z=1$ is the highest one.
In zeroth order, the Kramers pair of the highest confined valence subband $\mathrm{vb}_0$ therefore reads
\begin{equation}
    \ket{\mathrm{vb}_0^{(0)}\Uparrow}
    \equiv
    \ket{S_+ \uparrow;1,1}, \qquad
    \ket{\mathrm{vb}_0^{(0)}\Downarrow}
    \equiv
    \ket{S_- \downarrow;1,1},
    \label{eq:unperturbed_top_vb_state}
\end{equation}
where $\ket{S_\pm} = \left(\ket{S_x} \pm i\ket{S_y}\right)/\sqrt{2}$; see Table~\ref{tab:basis}.

\subsection{First-order admixture of $\mathrm{CB}_2$}

The linear-in-momentum coupling between the top valence band and the second conduction band admixes $\mathrm{CB}_2$ into $\mathrm{vb}_0$ with an amplitude of order $1/L$.
We first consider the spin-up partner.
At $k_x=0$, the relevant coupling in Eq.~\eqref{eq:hamiltonian} is
\begin{equation}
    \bra{Z\uparrow} H_{\mathbf{k}\cdot\mathbf{p}}
    \ket{S_+ \uparrow} = -i\frac{\hbar p_\perp}{\sqrt{2}m_\mathrm{e}}\, k_y .
    \label{eq:vb1_cb2_coupling}
\end{equation}
Since translational symmetry is broken along $y$, we replace $k_y\to -i\partial_y$.
First-order perturbation theory then gives
\begin{equation}\label{eq:vb_0_first_order_correction}
    \delta \ket{\mathrm{vb}_0^{(1)} \Uparrow}
    = \frac{\hbar p_\perp}{\sqrt{2}m_\mathrm{e}} \sum_{n=1}^{\infty}
    \frac{\bra{\phi_n}\partial_y\ket{\phi_1}}{\Delta_{n1}(L)}
    \ket{Z\uparrow;n,1},
\end{equation}
with
\begin{equation}
    \Delta_{n1}(L)
    =
    E_{\mathrm{CB}_2;n,1}^{(0)}
    -
    E_{\mathrm{VB}_1;1,1}^{(0)}
    > 0 .
    \label{eq:finite_size_energy_gap}
\end{equation}
Only $n_z=1$ appears because the coupling in Eq.~\eqref{eq:vb1_cb2_coupling} does not involve $k_z$.
Because of the derivative, the matrix element in the numerator of Eq.~\eqref{eq:vb_0_first_order_correction} is already of order $1/L$.
To leading order we may therefore replace $\Delta_{n1}(L)$ by its $L\to\infty$ limit, which, according to Eq.~\eqref{eq:2Dbox_spectrum}, is the bulk gap between the second conduction band and the top valence band,
\begin{equation}
    \Delta_{\mathrm{CB}_2,\mathrm{VB}_1} = E_{\mathrm{CB}_2} - E_{\mathrm{VB}_1}.
\end{equation}
This replacement is not uniform in $n$.
The $n$-dependent part of $\Delta_{n1}(L)/\Delta_{\mathrm{CB}_2,\mathrm{VB}_1}-1$ equals $(n\pi\ell/L)^2$, with $\ell = \hbar/\sqrt{2m^*_{\mathrm{CB}_2,y}\Delta_{\mathrm{CB}_2,\mathrm{VB}_1}}$, and is not small for $n \gtrsim L/\ell$.
We return to the contribution of these large-$n$ terms at the end of this appendix, after evaluating the Zeeman term, and show that it does not affect the bulk result.

Using the completeness of $\{\phi_n\}$, the correction in Eq.~\eqref{eq:vb_0_first_order_correction} becomes
\begin{align}\label{eq:vb0_upspin_correction}
    \delta \ket{\mathrm{vb}_0^{(1)} \Uparrow}
    &= \frac{\hbar p_\perp}{\sqrt{2}m_\mathrm{e}\Delta_{\mathrm{CB}_2,\mathrm{VB}_1}}
    \sum_{n=1}^{\infty} \bra{\phi_n}\partial_y\ket{\phi_1} \ket{Z\uparrow;n,1}
    \nonumber\\
    &= \frac{\hbar p_\perp}{\sqrt{2}m_\mathrm{e}\Delta_{\mathrm{CB}_2,\mathrm{VB}_1}}\,
    \phi'_1(y)\,\phi_1(z)\ket{Z\uparrow},
\end{align}
where $\phi'_1$ denotes the derivative of $\phi_1$.

An analogous derivation, or time reversal, gives the correction of the spin-down partner,
\begin{equation} \label{eq:vb0_downspin_correction}
    \delta \ket{\mathrm{vb}_0^{(1)} \Downarrow}
    = \frac{\hbar p_\perp}{\sqrt{2}m_\mathrm{e}\Delta_{\mathrm{CB}_2,\mathrm{VB}_1}}\,
    \phi'_1(y)\,\phi_1(z)\ket{Z\downarrow}.
\end{equation}
We write the first-order states as
\begin{equation}
    \ket{\mathrm{vb}_0^{(1)}\sigma} = \ket{\mathrm{vb}_0^{(0)}\sigma} + \delta\ket{\mathrm{vb}_0^{(1)}\sigma},
    \qquad \sigma \in \{\Uparrow, \Downarrow\};
\end{equation}
their norm is $1+\mathcal{O}(L^{-2})$.

\subsection{Zeeman splitting}

We choose the gauge
\begin{equation}
    \mathbf{A} = (-B_z y,0,0),
\end{equation}
which preserves translational invariance along the wire and has its origin at the wire center.
For an electron with charge $-e$, minimal coupling gives
\begin{equation}
    k_x
    \rightarrow
    k_x+\frac{e}{\hbar}A_x
    =
    k_x-\frac{eB_z y}{\hbar}.
    \label{eq:minimal_coupling}
\end{equation}
The term linear in $B_z$ produces the interband matrix element
\begin{equation}
    \bra{Z\uparrow}
    H_B^{\mathrm{orb}}
    \ket{S_+ \uparrow}
    =
    \frac{eB_zp_\perp}{\sqrt{2}m_\mathrm{e}}\,y .
    \label{eq:orbital_magnetic_matrix_element}
\end{equation}
The admixture is of order $1/L$, while the orbital vertex is proportional to $y$ and hence of order $L$; their product gives an $L$-independent contribution.
Combining Eq.~\eqref{eq:orbital_magnetic_matrix_element} with the $\mathrm{CB}_2$ admixture in Eqs.~\eqref{eq:vb0_upspin_correction} and \eqref{eq:vb0_downspin_correction}, and adding the spin Zeeman term of Eq.~\eqref{eq:spin_zeeman}, we obtain the matrix elements of the magnetic Hamiltonian $H_B$ in the $\mathrm{vb}_0$ Kramers pair:
\begin{align}
    \bra{\mathrm{vb}_0^{(1)}\Uparrow}H_B\ket{\mathrm{vb}_0^{(1)}\Uparrow}
    &= -\bra{\mathrm{vb}_0^{(1)}\Downarrow}H_B\ket{\mathrm{vb}_0^{(1)}\Downarrow}\nonumber\\
    &= \frac{\mu_\mathrm{B}}{2}\left(g_0 - \frac{2p_\perp^2}{m_\mathrm{e}\Delta_{\mathrm{CB}_2,\mathrm{VB}_1}}\right)B_z,
    \label{eq:Zeemanzz}\\
    \bra{\mathrm{vb}_0^{(1)}\Uparrow}H_B\ket{\mathrm{vb}_0^{(1)}\Downarrow}
    &= \bra{\mathrm{vb}_0^{(1)}\Downarrow}H_B\ket{\mathrm{vb}_0^{(1)}\Uparrow} = 0,
    \label{eq:Zeemanzz_offdiag}
\end{align}
where $\mu_\mathrm{B}=e\hbar/2m_\mathrm{e}$ and $g_0\simeq 2$ is the bare electron spin $g$ factor.
The off-diagonal elements vanish because neither the spin Zeeman term nor the orbital coupling in Eq.~\eqref{eq:orbital_magnetic_matrix_element} flips the spin.
In deriving Eq.~\eqref{eq:Zeemanzz}, we used
\begin{equation}
    \bra{\phi_1}y\,\partial_y\ket{\phi_1} = -\frac{1}{2},
    \label{eq:y_derivative_direct_evaluation}
\end{equation}
which holds for any real envelope that vanishes at the boundary: integration by parts gives $\bra{\phi_1}y\partial_y\ket{\phi_1} = -\bra{\phi_1}\partial_y y\ket{\phi_1}$, and $[\partial_y, y]=1$.
The result is therefore independent of $L$ and of the shape of the confining potential, as expected for a bulk quantity.

From Eq.~\eqref{eq:Zeemanzz}, the $g_{zz}$ element of the bulk $g$ tensor is
\begin{equation}
    g_{zz}
    =
    g_0
    -
    \frac{2p_\perp^2}
    {m_\mathrm{e}\Delta_{\mathrm{CB}_2,\mathrm{VB}_1}}
    \simeq
    -18.08,
    \label{eq:gzz_bulk_result}
\end{equation}
where we have used the parameter values in Table~\ref{tab:kp_parameters} for the numerical value.

\subsection{Finite-size corrections}

We now return to the replacement $\Delta_{n1}(L)\to\Delta_{\mathrm{CB}_2,\mathrm{VB}_1}$ in Eq.~\eqref{eq:vb_0_first_order_correction}.
We absorb the $n$-independent part of $\Delta_{n1}(L)-\Delta_{\mathrm{CB}_2,\mathrm{VB}_1}$ into the gap, since it only rescales the result by $1+\mathcal{O}(L^{-2})$.
Keeping the $n$-dependence of the denominators, the envelope $\phi_1'(y)$ in Eq.~\eqref{eq:vb0_upspin_correction} is replaced by
\begin{equation}
    \chi(y) = \sum_{n}\frac{\bra{\phi_n}\partial_y\ket{\phi_1}}{1+(n\pi\ell/L)^2}\,\phi_n(y).
\end{equation}
Since $(1-\ell^2\partial_y^2)\phi_n=[1+(n\pi\ell/L)^2]\phi_n$, the function $\chi$ is the solution of $(1-\ell^2\partial_y^2)\chi=\phi_1'$ with $\chi(\pm L/2)=0$:
\begin{equation}
    \chi(y) = \frac{1}{1+(\pi\ell/L)^2}
    \left[\phi_1'(y) - \phi_1'(L/2)\,\frac{\sinh(y/\ell)}{\sinh\left(L/2\ell\right)}\right].
    \label{eq:boundary_layer}
\end{equation}
The second term is a boundary layer of width $\ell$ that restores the hard-wall boundary condition.
Correspondingly, the factor $\bra{\phi_1}y\,\partial_y\ket{\phi_1}=-1/2$ in the orbital contribution to Eq.~\eqref{eq:Zeemanzz} is replaced by $\bra{\phi_1}y\ket{\chi}$.
This integral is elementary; and reads as
\begin{multline}
    \bra{\phi_1}y\ket{\chi}
    = \frac{1}{1+\pi^2\varepsilon^2}
    \left[
    -\frac{1}{2}
    + \frac{2\pi^2\varepsilon^2}{1+\pi^2\varepsilon^2}\right. \nonumber\\
    \left.- \frac{8\pi^2\varepsilon^3\coth\left(1/2\varepsilon\right)}{\left(1+\pi^2\varepsilon^2\right)^2}
    \right],
    \label{eq:orbital_closed_form}
\end{multline}
where $\varepsilon = \ell/L$.
In the limit $\varepsilon\to 0$ and substituting $\ell/L$ back, this reduces to 
\begin{equation}
    \bra{\phi_1}y\ket{\chi}
    = -\frac{1}{2}
    + \frac{5\pi^2}{2}\left(\frac{\ell}{L}\right)^2
    - 8\pi^2\left(\frac{\ell}{L}\right)^3
    + \mathcal{O}\!\left(\left(\frac{\ell}{L}\right)^4\right).
\end{equation}
This recovers the bulk limit, and shows that the correction from the denominator is only second or higher order.
 
Now we show that by symmetry, there is no $1/L$ correction to the $g$ tensor.
Consider any term in the perturbative expansion of the matrix elements in Eqs.~\eqref{eq:Zeemanzz} and \eqref{eq:Zeemanzz_offdiag}. Each factor $k_y$ or $k_z$ in the correction of the wavefunction contributes as a factor of $L^{-1}$ and flips the parity of the envelope under $y\to-y$ or $z\to-z$, respectively. Each factor $y$ from minimal coupling contributes $L^{+1}$ and flips the parity under $y\to-y$. Band energies, spin–orbit, and spin Zeeman terms carry neither a power of $L$ nor a parity flip, and the energy denominators depend on $L$ only through $L^{-2}$ [Eq.~\eqref{eq:2Dbox_spectrum}]. Thus the power of $L$ of every term has the same parity as its total number of parity flips. Since the initial and final envelopes, $\phi_1(y)\phi_1(z)$, are even in both $y$ and $z$, the number of flips must be even. Since the expansion of the energy denominators in powers of $L^{-2}$ produces only even powers of $L$, and in particular no $L^{-1}$ term. This argument also covers the B-linear terms that arise from second-order elements such as $k_xk_y\to -(eB_z/\hbar)\{y,k_y\}/2$, which are even under $y\to-y$. Because this expansion is not uniform in $n$, the resummed large-$n$ terms discussed above also produce odd powers, but these start at $L^{-3}$.
 
The leading finite-size correction is therefore of order $L^{-2}$. However, it cannot be calculated consistently within a second-order $\mathbf{k}\cdot\mathbf{p}$ model, because third-order-in-momentum terms contribute at the same order: they correct the wavefunction at order $L^{-3}$ and hence the $g$ factor at order $L^{-2}$.

\section{Quadratic Stark effect}
\label{app:quadratic_stark_effect}

In this appendix, we derive the quadratic Stark shift of the top confined valence subband of the nanowire caused by a uniform electric field
$\mathcal{E}_y$ applied along the $y$ direction, perpendicular to the wire axis $x$. We use a single-band effective-mass model, in which each state is written as the product of the band-edge Bloch function $|v\rangle$ and a slowly varying envelope function. The coupling to the other bands is absorbed into the anisotropic effective masses $m_y^*$ and $m_z^*$ along the two confinement directions,
calculated in Appendix~\ref{app:effmass}. This description is justified when the other bands are energetically well separated from the top valence band on
the scale of the confinement and Stark energies. The envelope function then obeys a Schr\"odinger equation with these effective masses. 
Moreover, the Kramers degeneracy is omitted here, because it is not effected at the $\Gamma$ point by the electric field.

We model the nanowire cross-section as a square of side $L$ with hard walls and treat the field in second-order nondegenerate perturbation theory, which is valid as long as $e\mathcal{E}_yL$ is small compared with the spacing $E_{2,1}-E_{1,1}$ between the lowest transverse subbands. Throughout, confinement energies are given in the hole picture, i.e., measured positive downward from the valence-band edge.

The confinement is described by a two-dimensional infinite square-well
potential in the $(y,z)$ plane,
\begin{align}
V(y,z) =
\begin{cases}
0, & y,z \in [-L/2, L/2], \\
\infty, & \text{otherwise}.
\end{cases}
\end{align}
The envelope eigenstates factorize as
$\psi_{n_y,n_z}(y,z) = \phi_{n_y}(y)\,\phi_{n_z}(z)$, with
\begin{align}
\phi_n(u)
= \sqrt{\frac{2}{L}}\,
\sin\!\left[\frac{n\pi}{L}\left(u+\frac{L}{2}\right)\right],
\qquad u = y, z,
\end{align}
where $n_y,n_z=1,2,3,\dots$ label the transverse modes along the $y$ and $z$
directions, respectively. The corresponding envelope energies are
\begin{align}
E_{n_y,n_z}
= \frac{\pi^2\hbar^2}{2m_\mathrm{e}L^2}
\left(
\frac{n_y^2}{m_y^*}
+
\frac{n_z^2}{m_z^*}
\right),
\end{align}
where $m_y^*$ and $m_z^*$ are the dimensionless effective masses along the
$y$ and $z$ directions, expressed in units of the free-electron mass
$m_\mathrm{e}$.

We denote the envelope state $\psi_{n_y,n_z}$ by
$|n_y,n_z\rangle = |n_y\rangle \otimes |n_z\rangle$, where $|n_y\rangle$
corresponds to $\phi_{n_y}(y)$. The top confined valence subband is then
\begin{align}
|\mathrm{vb}_0\rangle
= |1,1\rangle \otimes |v\rangle,
\end{align}
where $|1,1\rangle$ is the transverse envelope ground state and $|v\rangle$
is the top valence-band Bloch function.

The uniform electric field $\mathcal{E}_y$ along the $y$ direction enters
through
\begin{align}
H_{\mathcal{E}} = e \mathcal{E}_y y,
\end{align}
where $e>0$ is the elementary charge and $y$ is measured from the center of the cross-section. Consistent with the single-band description, we neglect interband
matrix elements of $y$, so that $H_{\mathcal{E}}$ acts only on the envelope. Because it depends only on $y$, it couples $|1,1\rangle$ only to the states
$|n_y,1\rangle$, leaving the Bloch function and the $z$ envelope unaffected. The first-order correction is proportional to $\langle 1|y|1\rangle$ and therefore vanishes
because $\phi_1^2(y)$ is even. 
Thus, the leading correction to the energy of the
envelope ground state is of second order,
\begin{align}
E^{(2)}
= e^2\mathcal{E}_y^2
\sum_{n_y \neq 1}
\frac{
\left|\langle n_y| y |1\rangle\right|^2
}{
E_{1,1}-E_{n_y,1}
}.
\end{align}

Parity restricts the sum: $\phi_1(y)$ is even and $y$ is odd, so $\langle n_y|y|1\rangle$ is nonzero only for odd envelope functions, i.e.,
those with even $n_y$. For even $n_y$ we obtain
\begin{align}
\langle n_y| y |1\rangle
&= \int_{-L/2}^{L/2} dy\,
\phi_{n_y}(y)\,y\,\phi_1(y)
\nonumber \\
&=
-\frac{8Ln_y}{\pi^2(n_y^2-1)^2}.
\end{align}
The energy denominator is
\begin{align}
E_{1,1}-E_{n_y,1}
=
-\frac{\pi^2\hbar^2}{2m_\mathrm{e}m_y^*L^2}
(n_y^2-1).
\end{align}
Hence, the second-order correction becomes 
\begin{align}
E^{(2)}
&=
-\frac{128\,m_\mathrm{e}m_y^* e^2 \mathcal{E}_y^2 L^4}
{\pi^6\hbar^2}
\sum_{n_y=2,4,6,\dots}
\frac{n_y^2}{(n_y^2-1)^5}
\nonumber \\
&=
\mathcal{C}\,m_y^*\mathcal{E}_y^2L^4.
\end{align}
The sum can be evaluated in closed form,
$\sum_{n_y=2,4,6,\dots} n_y^2/(n_y^2-1)^5 = \pi^2(15-\pi^2)/3072 \approx 0.016483$. It is dominated by the $n_y=2$ term, $4/243$, which alone accounts for $99.9\%$ of the total, so the shift is essentially that of the two-level system formed by the two lowest $y$ subbands. This gives 
\begin{align}
\mathcal{C}
&=
-\frac{128\,m_\mathrm{e} e^2}{\pi^6\hbar^2}
\sum_{n_y=2,4,6,\dots}
\frac{n_y^2}{(n_y^2-1)^5}
\nonumber \\
&=
-\frac{15-\pi^2}{24\pi^4}\,
\frac{m_\mathrm{e} e^2}{\hbar^2}
\approx
-2.1945\times10^{-3}\,
\frac{m_\mathrm{e} e^2}{\hbar^2}.
\label{eq:c_stark}
\end{align}
In practical units,
$\mathcal{C} \approx -2.88\times10^{-7}\,
\mathrm{meV}\,(\mathrm{kV/cm})^{-2}\,\mathrm{nm}^{-4}$.
Because the energies are given in the hole picture, $E^{(2)}<0$ corresponds to an upward shift of the top valence subband in electron energy.

\section{Projection of nanowire states onto bulk-band components}
\label{app:overlap}

In this appendix, we define a measure of the bulk-band character of the nanowire eigenstates at $k_x=0$. We do this by projecting each state, site by site, onto the bulk eigenvectors at the $\Gamma$ point, which gives the weight $W_n^M$ of the bulk-band manifold $M$ in the nanowire state $n$.

In the discretized ten-band tight-binding model used in the numerical calculations, the transverse cross section is represented by a set of lattice sites $(y,z)$, each carrying a ten-dimensional internal degree of freedom $a=1,\dots,10$ that comprises the orbital and spin components. The nanowire eigenstates at $k_x=0$ are written as
\begin{align}
    |\psi_n\rangle = \sum_{y,z} \sum_{a=1}^{10} \psi^{(n)}_{a}(y,z)\, |y,z;a\rangle ,
\end{align}
where $|y,z;a\rangle$ denotes the internal state $a$ on site $(y,z)$, and the amplitudes are normalized as $\sum_{y,z}\sum_{a}|\psi^{(n)}_{a}(y,z)|^2=1$. We denote by $\Psi^{(n)}(y,z)$ the ten-component local spinor with entries $\psi^{(n)}_{a}(y,z)$.

The bulk reference states are obtained by diagonalizing Eq.~\eqref{eq:hamiltonian} at $\mathbf{k}=0$. They coincide with the basis states of Table~\ref{tab:basis}, except for the second and third valence bands, which are hybridized by the off-diagonal spin-orbit term $\sqrt{2}\Delta_3$. We denote these eigenstates by $u^{(M,s)}$, where $M$ labels the five Kramers-degenerate manifolds $\mathrm{VB}_1$, $\mathrm{VB}_2$, $\mathrm{VB}_3$, $\mathrm{CB}_1$, and $\mathrm{CB}_2$, and $s\in\{\uparrow,\downarrow\}$ labels the two Kramers components. Together, the ten vectors $u^{(M,s)}$ form an orthonormal basis of the internal space.

We define the weight of the bulk-band manifold $M$ in the nanowire state $n$ as
\begin{align}
    W_n^M = \sum_{y,z} \sum_{s\in\{\uparrow,\downarrow\}} \left| \sum_{a=1}^{10} \left[u^{(M,s)}_{a}\right]^* \psi^{(n)}_{a}(y,z) \right|^2 \\
    \label{eq:bulk_weight}
\end{align}
Completeness of the bulk basis implies the sum rule
\begin{align}
    \sum_M W_n^M = 1 .
\end{align}
The weight $W_n^M$ equals unity if and only if the local spinor on every site lies within a single band. This is the discrete counterpart of the envelope-function picture, in which a confined state is the product of a transverse envelope and a bulk band-edge state. Deviations from unity arise because a confined state also contains bulk components with nonzero transverse wave vectors, whose eigenvectors differ from those at $\Gamma$. For band-edge states of wires whose cross section is large compared with the lattice spacing, the envelope varies slowly from site to site and these components are small. $W_n^M$ then approaches unity for a single band, unless the confinement energy is comparable to the splitting between bulk manifolds.  

\section{Symmetry quantum numbers of the dominant intermediate subbands}
\label{app:dominant_subband}

The intermediate Kramers pairs responsible for the dominant Rashba contributions are identified in Sec.~\ref{symmetry_analysis}. Here, we explain how their symmetry quantum numbers in Table~\ref{tab:level_symmetry} are assigned and relate these assignments to their transverse envelope structure
and bulk-band character. 

At $k_x=0$, the confined nanowire eigenstates can be classified by the commuting symmetry operations $C_{2z}$ and $\mathcal{I}$. The corresponding symmetry labels are defined through Eq.~\eqref{eq:symmetry_labels}.

\begin{figure}[t]
    \centering
    \includegraphics[width=1\columnwidth]{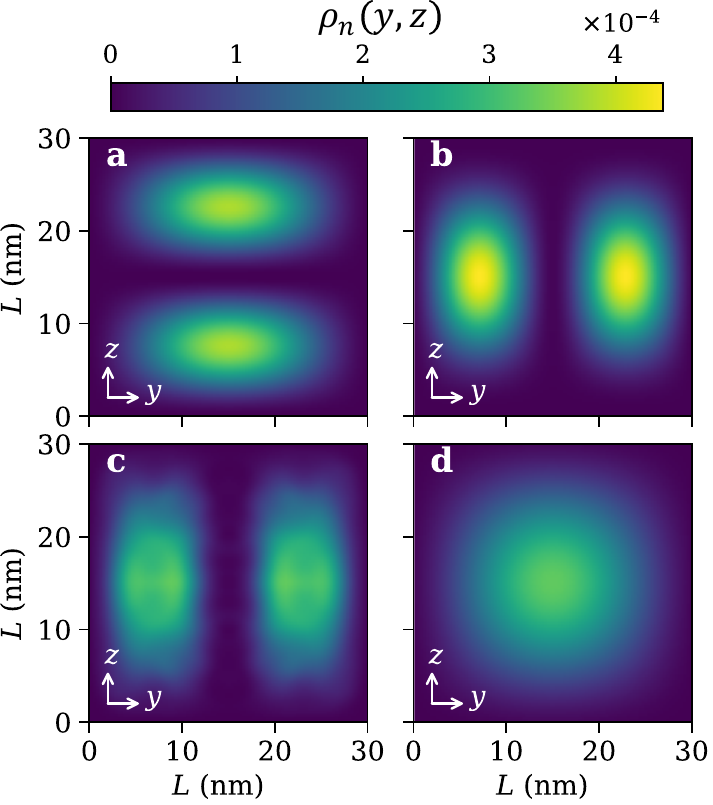}

\caption{
Real-space probability distributions of the unique dominant intermediate subband states contributing to the Rashba splitting for a nanowire with cross-sectional side length $L=30~\mathrm{nm}$. (a) The near-valence state that contributes dominantly for a $z$-directed electric field, $\mathcal{E}\parallel z$. (b) and (c) The valence band states that contribute for both $\mathcal{E}\parallel y$ and $\mathcal{E}\parallel z$. (d) The conduction band state that contributes mainly for $\mathcal{E}\parallel y$. The plotted quantity is the transverse probability distribution
$\rho_n(y,z)=\sum_{a=1}^{10}|\psi_a^{(n)}(y,z)|^2$
for one representative member of each Kramers pair. }

    \label{fig:important_states}
\end{figure}

It is useful to separate the total symmetry label into an internal and an envelope contribution,
\begin{align}
    \lambda_g^{\mathrm{tot}}
    =
    \lambda_g^{\mathrm{int}}
    \lambda_g^{\mathrm{env}},
    \qquad
    g=C_{2z},\mathcal{I}.
    \label{eq:total_symmetry_product}
\end{align}
Here $\lambda_g^{\mathrm{int}}$ is determined by the dominant bulk-band spinor component, while $\lambda_g^{\mathrm{env}}$ is determined by the parity of the confined envelope wave function. This distinction is important because the probability density $\rho_n(y,z)=\sum_{a=1}^{10}|\psi_a^{(n)}(y,z)|^2$ does not contain the sign or phase information of the wave function. Therefore, a state with an apparently symmetric density can still be odd under inversion if its dominant internal orbital component is inversion odd. The density plots in Fig. \ref{fig:important_states} should therefore be used only as a guide to the dominant envelope character; the final symmetry assignment is made using the full ten-component spinor.

For an ideal square cross section, the envelope functions can be approximately classified by box-like modes
\begin{align}
\begin{split}
    \phi_{n_y,n_z}(y,z)
    &\sim
    \sin\left[\frac{n_y\pi}{L}\left(y+\frac{L}{2}\right)\right]\\
    &\times \sin\left[\frac{n_z\pi}{L}\left(z+\frac{L}{2}\right)\right] 
\end{split}
\end{align}
With respect to the center of the wire, the envelope parity along $y$ is even for odd $n_y$ and odd for even $n_y$; similarly, the envelope parity along $z$ is even for odd $n_z$ and odd for even $n_z$. Thus, the lowest envelope $(n_y,n_z)=(1,1)$ is even in both directions, the mode $(2,1)$ is odd in $y$ and even in $z$, and the mode $(1,2)$ is even in $y$ and odd in $z$.

Fig. \ref{fig:important_states}(a) has a clear two-lobe structure along the $z$ direction and is well described by a dominant $(n_y,n_z)=(1,2)$-like envelope. This envelope is even under $C_{2z}$ but odd under inversion. Since the state is mainly VB$_1$-like in its internal character, the odd envelope parity under inversion changes the total inversion eigenvalue relative to the top vb$_0$ state while preserving the $C_{2z}$ branch. Therefore, this state has the symmetry assignment
\begin{align}
    \ket{\uparrow}\text{ branch}: (+i,-1),
    \qquad
    \ket{\downarrow}\text{ branch}: (-i,-1).
\end{align}

Figs.~\ref{fig:important_states}(b) and \ref{fig:important_states}(c) show states with a dominant two-lobe structure along the $y$ direction. They are therefore approximately described by $(n_y,n_z)=(2,1)$-like envelopes, which are odd in $y$ and even in $z$. Such an envelope flips both the $C_{2z}$ and inversion quantum numbers relative to a VB$_1$-like internal component.
Consequently, these states belong to the symmetry sectors
\begin{align}
    \ket{\uparrow}\text{ branch}: (-i,-1),
    \qquad
    \ket{\downarrow}\text{ branch}:(+i,-1).
\end{align}

Although the states in Figs.~\ref{fig:important_states}(b) and \ref{fig:important_states}(c) have the same dominant envelope parity and hence the same total symmetry quantum numbers, their microscopic orbital composition is different. From the bulk-band projection in Fig.~\ref{fig:overlap}(c), the state in Fig.~\ref{fig:important_states}(b) is mainly derived from the top bulk valence band, VB$_1$, whereas the state in Fig.~\ref{fig:important_states}(c) has dominant contribution from the second valence bulk band, VB$_2$. Therefore, their different orbital composition and energy separation lead to different quantitative contributions to the Rashba splitting, even though they satisfy the same symmetry selection rule.

Finally, Fig.~\ref{fig:important_states}(d) shows the dominant conduction-sector intermediate state. Its density is approximately circular, indicating an even envelope in both transverse directions. In this case, the total symmetry is mainly determined by the internal conduction-band spinor character. The bulk-band projection shows that this state is predominantly derived from the second conduction bulk band, i.e., it has dominant CB$_2$ character. The resulting symmetry assignment is
\begin{align}
    \ket{\uparrow}\text{ branch}: (-i,-1),\qquad
    \ket{\downarrow}\text{ branch}: (+i,-1),
\end{align}
which again allows coupling to the top vb$_0$ doublet through the $y$-dipole operator and through $V_1$.

\bibliography{ref}
\end{document}